\PassOptionsToPackage{monochrome}{color}
\PassOptionsToPackage{monochrome}{xcolor}

\documentclass[11pt]{IEEEojcsys}

\usepackage{color,array}

\usepackage{graphicx}
 \jvol{}
 \jnum{}
 \paper{}
 \pubyear{}
 \receiveddate{}
 \date{}
 \accepteddate{}
 \publisheddate{}
 \currentdate{}
 \doiinfo{}

\newtheorem{theorem}{Theorem}
\newtheorem{lemma}{Lemma}
\newtheorem{proposition}{Proposition}
\usepackage{amsmath,amssymb}
\usepackage{cite}
\usepackage{microtype}
\usepackage{xcolor}
\usepackage{braket}
\usepackage[colorlinks=true,linkcolor=teal,citecolor=teal,urlcolor=teal]{hyperref}

\newtheorem{definition}{Definition}

\newtheorem{corollary}{Corollary}

\newcommand{\BPP}{\mathsf{BPP}}
\newcommand{\BQP}{\ensuremath{\mathsf{BQP}}}

\def\defas{\ensuremath{\mathrel{:=}}}

\DeclareMathDelimiter{\orbrack}{\mathopen}{operators}{"5D}{largesymbols}{"03}
\DeclareMathDelimiter{\clbrack}{\mathclose}{operators}{"5B}{largesymbols}{"02}
\def\intcc#1{\ensuremath{[#1]}}

\usepackage{etoolbox}

\makeatletter
\patchcmd{\ps@headings}
  {\includegraphics{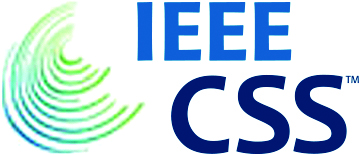}}
  {}
  {}
  {\PackageError{customstyle}{Cannot remove running-page logo}{}}

\patchcmd{\ps@plain}
  {\includegraphics{ojcsys-logo}}
  {}
  {}
  {\PackageError{customstyle}{Cannot remove first-page logo}{}}
\makeatother

\AtBeginDocument{%
  \definecolor{ojcolor}{rgb}{0,0,0}%
  \hypersetup{hidelinks}%
  \pagestyle{headings}%
}

\makeatletter
\patchcmd{\ps@plain}
  {\hbox to \textwidth{\vrule width42pc height.75pt depth0pt}}
  {}
  {}
  {\PackageWarning{customstyle}{First-page rule not found}}
\makeatother

\usepackage{cleveref}
\crefname{theorem}{Theorem}{Theorems}
\crefname{lemma}{Lemma}{Lemmas}
\crefname{section}{Section}{Sections}
\crefname{proposition}{Proposition}{Propositions}
\crefname{appendix}{Appendix}{Appendices}

\begin{document}

\title{Succinct Infinite-Horizon Lyapunov and LQR Value Problems Are
BQP-Complete}

\author{Mohamed Serry\affilmark{1,2}}

\author{Jun Liu\affilmark{1}}%

\affil{Department of Applied Mathematics,
University of Waterloo, 200 University Avenue West, Waterloo, ON N2L 3G1,
Canada}

\affil{Department of Mechanical and
Mechatronics Engineering, University of Waterloo, 200 University Avenue West, Waterloo, ON N2L 3G1,
Canada}

\authornote{Both authors contributed equally to this manuscript. This work was supported by the National Research Council of Canada (NRC) through the
Applied Quantum Computing (AQC) Challenge program under project AQC-318. J.L.
also acknowledges the support of the Natural Sciences and Engineering Research Council
of Canada (NSERC) and the Canada Research Chairs program.}
\begin{abstract}
Fundamental tasks in linear systems analysis and control require evaluating infinite-horizon output energies for autonomous systems and optimal value functions for controlled systems. These quantities are characterized by Lyapunov and algebraic Riccati equations, respectively. In this work, we study promise problems for their evaluation when the system matrices are sparse and represented succinctly by Boolean circuits specifying their nonzero entries, with input length polynomial in the logarithm of the system dimension. For dissipative autonomous systems with coordinate-vector initial states, we define the succinct Lyapunov value problem (SLV). Its extension to linear-quadratic optimal control is the succinct LQR value problem (SLQRV). We prove that SLV is $\BQP$-hard: every problem in bounded-error quantum polynomial time ($\BQP$) reduces to it in polynomial time. We then establish that the more general SLQRV problem admits a quantum algorithm running in time polynomial in the succinct input length. Since SLV is a special case of SLQRV, both problems are $\BQP$-complete. Consequently, unless bounded-error probabilistic polynomial time coincides with $\BQP$, no classical randomized algorithm can solve either problem on all promised instances in time polynomial in the succinct input length.
\end{abstract}
\maketitle
\section{Introduction}
\label{sec:Introduction}
In systems analysis and control, Lyapunov and algebraic Riccati equations provide fundamental links between linear dynamics and infinite-horizon quadratic performance measures. Lyapunov equations characterize the output energy of stable autonomous systems and the cost of fixed stabilizing feedback policies, whereas Riccati equations characterize the optimal value function of linear-quadratic regulators (LQRs). These connections underpin many methods for performance analysis and control design, including policy-gradient methods \cite{mohammadi2021convergence,clayton2024}. 

In this work, we study the computational complexity of evaluating both output energies and optimal LQR costs when the system matrices are specified succinctly.

Consider first the real autonomous linear system
\[
    \dot{x}(t)=Ax(t),\qquad y(t)=Cx(t),\qquad x(0)=b,
\]
where \(x(t)\in\mathbb{R}^n\), \(y(t)\in\mathbb{R}^m\), \(A\in\mathbb{R}^{n\times n}\) is Hurwitz, \(C\in\mathbb{R}^{m\times n}\), and \(b\in\mathbb{R}^n\). Its infinite-horizon output energy is
\begin{equation}\label{eq:Energy}
        J(b)\defas\int_0^\infty y(t)^\intercal y(t)\,\mathrm{d}t
    =b^\intercal Pb,
\end{equation}
where the observability Gramian
\[
    P\defas\int_0^\infty
    e^{A^\intercal t}C^\intercal Ce^{At}\,\mathrm{d}t
\]
is the unique symmetric positive-semidefinite solution of the Lyapunov equation
\begin{equation}\label{eq:Lyapunov}
    A^\intercal P+PA+C^\intercal C=0.
\end{equation}
The scalar \(J(b)\) also equals the squared \(H_2\) norm of  $ H(\zeta)=C(\zeta I-A)^{-1}b$, viewed as the transfer function from a scalar input to \(y\) \cite{zhou1996}. Thus, \(J(b)\) measures the impulse-response energy associated with the input direction \(b\).

For controlled systems, consider
\[
    \dot{x}(t)=Ax(t)+Bu(t),\qquad x(0)=b,
\]
where \(B\in\mathbb{R}^{n\times p}\) and \(u(t)\in\mathbb{R}^p\). Given \(C\in\mathbb{R}^{m\times n}\) and \(R=R^\intercal\succ0\), the infinite-horizon LQR optimal value function is
\begin{equation}
\label{eq:ValueFunction}
    V(b)\defas\inf_{u(\cdot)\in L^{2}}
    \int_0^\infty
    \left(\|Cx(t)\|^2+u(t)^\intercal Ru(t)\right)\,\mathrm{d}t.
\end{equation}
Under the standard stabilizability and detectability assumptions,
    $V(b)=b^\intercal P_\star b$, where \(P_\star\succeq0\) is the unique stabilizing solution of the algebraic Riccati equation
\begin{equation}\label{eq:Riccati} 
\begin{split}
    A^\intercal P_\star&+P_\star A
    -P_\star G  P_\star+Q=0,\\
    G&\defas BR^{-1}B^\intercal,\\
  Q& \defas C^{\intercal}C.  
\end{split}
\end{equation}
The corresponding optimal feedback is
\(u=-R^{-1}B^\intercal P_\star x\).

The two value-evaluation tasks are closely related. For a fixed stabilizing feedback \(u=-Kx\), the LQR cost is characterized by the Lyapunov equation
\[
    (A-BK)^\intercal P_K+P_K(A-BK)
    +C^\intercal C+K^\intercal RK=0,
\]
and equals \(b^\intercal P_Kb\). Such policy evaluations arise, for example, in policy-gradient methods \cite{mohammadi2021convergence,clayton2024}.

The importance of Lyapunov and Riccati equations in stability analysis, performance evaluation, model reduction, and optimal control has motivated extensive research on their numerical solution. Classical approaches include the Bartels--Stewart method for Lyapunov and Sylvester equations, Schur methods for algebraic Riccati equations, and iterative methods based on Newton--Kleinman iteration \cite{bartels1972,laub1979,benner2008}. For large sparse systems, low-rank alternating-direction implicit and Krylov-subspace methods exploit additional structure to reduce storage and computational costs \cite{benner2008,simoncini2016}. These developments make many large-scale problems computationally tractable, with complexity polynomial in the system dimension. Their efficiency, however, depends on properties such as spectral behavior, numerical rank, and the availability of effective approximations. Efficient evaluation of performance measures for systems of substantially higher dimension therefore remains an important computational objective.

The potential of quantum computers to accelerate numerical linear algebra has motivated quantum algorithms for Lyapunov and Riccati equations and their associated control problems. Quantum linear-system solvers and quantum singular value transformation provide foundational tools for representing and processing matrix solutions without explicitly computing all their entries \cite{hhl2009,gilyen2019}. Sun and Zhang proposed an early quantum algorithm for Lyapunov equations \cite{sun2017}, while Clayton et al.\ developed a block-encoding of the Lyapunov solution and used it to accelerate cost and gradient estimation within an LQR policy-gradient method \cite{clayton2024}. Subsequent developments include probabilistic preparation of mixed-state representations of Lyapunov solutions \cite{benedetti2026} and block-encoding algorithms for general linear matrix equations \cite{somma2026}. For Riccati equations, quantum approaches include matrix-geometric-mean constructions for particular classes of equations \cite{liu2025} and spectral-projector methods for continuous- and discrete-time algebraic and differential Riccati problems \cite{wang2026near}. Complementing these matrix-solution approaches, Huang et al.\ formulate finite-horizon LQR through state--costate linear systems to estimate optimal costs and control inputs directly \cite{slvLqrHuangFinite}.

The computational advantages of these algorithms depend on both the input representation and the information required from the solution. Preparing a quantum representation of a matrix does not, by itself, provide an efficient procedure for reading all its entries. Moreover, explicitly listing a sparse system matrix can already require storage proportional to the state dimension, while explicitly returning a dense Gramian or Riccati solution requires quadratically many entries. By contrast, many control applications require only a scalar performance measure, such as the output energy associated with an initial condition or the corresponding optimal LQR cost. These scalar quantities provide natural targets for quantum algorithms whose runtimes can depend only polylogarithmically on dimension, provided that the remaining complexity parameters are suitably bounded.

These observations also motivate a complexity-theoretic question: when quantum algorithms efficiently evaluate such quantities, can classical randomized algorithms achieve comparable guarantees under the same input representation? Algorithmic speedups over particular classical methods do not alone resolve this question. Establishing computational hardness requires showing that the evaluation task can encode arbitrary quantum computations.

We investigate this question using succinct representations of sparse system matrices, in which Boolean lookup circuits specify their nonzero entries \cite{papadimitriou1986,janzing2007}. The resulting descriptions have length polynomial in the logarithm of the relevant matrix dimensions. We consider two promise-decision problems: the \emph{Succinct Lyapunov Value} (SLV) problem, concerning infinite-horizon output energy, and the \emph{Succinct LQR Value} (SLQRV) problem, concerning optimal infinite-horizon quadratic cost. Both tasks require a decision about a scalar value rather than an explicit matrix solution. Their precise assumptions and promises are given below.

We prove that SLV is hard for promise bounded-error quantum polynomial time ($\BQP$), even for strictly dissipative autonomous systems whose system matrices have a single repeated real eigenvalue. We then utilize existing efficient quantum algorithms \cite{wang2026near} to establish that the broader SLQRV problem belongs to \(\BQP\) under quantitative stabilizability and detectability assumptions. We derive the required algorithmic bounds from these control-theoretic assumptions, without requiring a stabilizing feedback gain as input. Since SLV embeds into SLQRV with the same value and decision thresholds, these results establish \(\BQP\)-completeness of both problems. Consequently, unless \(\BPP=\BQP\), neither problem admits a classical randomized algorithm that solves all promised instances in time polynomial in the succinct input length.

Our results complement earlier complexity classifications for sparse linear systems \cite{hhl2009}, matrix powers \cite{janzing2007}, and matrix geometric means \cite{liu2025}. They also sharpen the connection between quantum complexity and control-performance evaluation. Somma et al.\ show that general linear matrix equations encompass \(\BQP\)-complete computations \cite{somma2026}; our SLV result establishes hardness within the more structured setting of infinite-horizon output energy for strictly dissipative autonomous systems. Wang et al.\ establish \(\BQP\)-completeness for selected values on explicit circuit-generated families of discrete-time algebraic and finite-horizon differential Riccati equations \cite{wang2026near}. Here, we classify continuous-time infinite-horizon output-energy and optimal-cost problems over the full promise classes defined in this work. Together, our hardness and membership results identify a conditional obstruction to efficient classical evaluation under the same succinct-input model used by the quantum algorithms.


\section{Preliminaries and Notation}
\label{sec:Preliminaries}

Let $\mathbb{Q}$, $\mathbb{R}$, $\mathbb{C}$, $\mathbb{R}_{+}$, $\mathbb{Z}$, and
$\mathbb{Z}_{+}$ denote the sets of rational numbers, real numbers, complex numbers,
nonnegative real numbers, integers, and nonnegative integers, respectively.
We define $\mathbb{N}\defas\mathbb{Z}_{+}\setminus\{0\}$. For
$a,b\in\mathbb{R}$ with $a\leq b$, 
$\intcc{a;b}\defas\intcc{a,b}\cap\mathbb{Z}$.    For functions $f,g:\mathbb{R}_{+}\to\mathbb{R}_{+}$, we write
$f(t)=O(g(t))$ if there exist constants $c>0$ and $t_0\geq 0$ such that
$f(t)\leq c g(t)$
     for all $t\geq t_0$.
Similarly, $f(t)=\Omega(g(t))$ if there exist constants $c>0$ and $t_0\geq
0$ such that
    $f(t)\geq c g(t)$
for all $t\geq t_{0}$. 
Finally, $f(t)=\Theta(g(t))$ if both $f(t)=O(g(t))$ and
$f(t)=\Omega(g(t))$. These definitions also apply to functions defined on
$\mathbb{N}$, with $t$ restricted to $\mathbb{N}$.

\subsection{Vectors and Matrices}
 The Euclidean dot product of  vectors $v,w\in \mathbb{C}^{d}$ is denoted  $\braket{v|w}$. Here, \(\|\cdot\|\) denotes the Euclidean norm for vectors and the induced operator \(2\)-norm for matrices. For vectors $v\in\mathbb{C}^{d_1}$ and $w\in\mathbb{C}^{d_2}$, their tensor
product $v\otimes w\in\mathbb{C}^{d_1d_2}$ is the vector whose
entries are $v_iw_j$, ordered lexicographically by $(i,j)$. The tensor product of matrices is defined analogously: for
\(M\in\mathbb{C}^{m_1\times n_1}\) and
\(N\in\mathbb{C}^{m_2\times n_2}\), the matrix
\(M\otimes N\in\mathbb{C}^{m_1m_2\times n_1n_2}\) has entries
\((M\otimes N)_{(i,j),(k,\ell)}=M_{ik}N_{j\ell}\), with row and column
index pairs ordered lexicographically. For a matrix $U\in \mathbb{C}^{d_{1}\times d_{2}}$,  \(U^\dagger\) denotes the conjugate transpose of \(U\). For matrices \(M\in\mathbb{C}^{p\times p}\) and
\(N\in\mathbb{C}^{q\times q}\), we define their  block-diagonal direct sum as  $ M\oplus N
    \defas
       \mathrm{blkdiag} (M,~N)
    \in\mathbb{C}^{(p+q)\times(p+q)}$. For real symmetric \(X,Y\in\mathbb{R}^{n\times n}\), \(X\succeq Y\) (\(X\succ Y\)) means \(X-Y\) is positive semidefinite (definite). A matrix \(U\in\mathbb{C}^{d\times d}\) is \emph{unitary} if   $U^\dagger U = UU^\dagger = I_d$ ($U^{-1}=U^\dagger$). A real matrix \(U\in\mathbb{R}^{d\times d}\) is \emph{orthogonal} if   $U^\intercal U = UU^\intercal = I_d$ ($U^{-1}=U^\intercal$). Matrix $S\in \mathbb{R}^{d\times d}$ is said to be \emph{$\gamma$-dissipative} if for some $\gamma>0$,
    \(-({S+S^\intercal})/{2}\succeq\gamma I\). The $\gamma$-dissipativity condition  implies $\|e^{St}\|\leq e^{-\gamma t}$,
and therefore guarantees asymptotic stability. 
Note that $\gamma$-dissipativity is strictly stronger than asymptotic stability. 

\subsection{Quantum Computing Notation}

We use standard Dirac notation for vectors in complex vector spaces. A
column vector $v\in\mathbb{C}^d$ is denoted by a \emph{ket}, $\ket{v}$, and
its conjugate transpose by a \emph{bra}, $\bra{v}$. Throughout,
a \emph{quantum state} means a pure quantum state, represented by a
unit-norm vector $\ket{\psi}\in\mathbb{C}^d$ ($\braket{\psi|\psi}=1$).   A \emph{qubit} is a quantum state in $\mathbb{C}^{2}$. Its standard computational basis vectors are \(\ket{0}\defas(1,0)^{\intercal}\) and \(\ket{1}\defas(0,1)^{\intercal}\), and we define the Hadamard basis vectors \(\ket{\pm}\defas(\ket{0}\pm\ket{1})/\sqrt{2}\).  The state space of a register of \(q\) qubits is the Hilbert space
\(\mathbb{C}^{2^q}\). For a bit string
\(z=z_1\cdots z_q\in\{0,1\}^q\), the corresponding computational-basis
vector is $\ket{z}
    \defas
    \ket{z_1}\otimes\cdots\otimes\ket{z_q}$.
When no ambiguity can arise, we may also label \(\ket{z}\) by the nonnegative integer whose \(q\)-bit binary representation is \(z\), and write this integer in its usual base-\(10\) representation.
 \emph{Ancillas} are auxiliary qubits used as temporary workspace during a quantum computation. They are typically initialized in a known state, such as \(\ket{0}\) for one ancilla or \(\ket{0}^{\otimes r}\) for \(r\) ancillas. For a single qubit, define the Pauli operators
\(X\defas\ket{0}\bra{1}+\ket{1}\bra{0}\),
\(Y\defas-i\ket{0}\bra{1}+i\ket{1}\bra{0}\), and
\(Z\defas\ket{0}\bra{0}-\ket{1}\bra{1}\). Note that
$X\ket{0}=\ket{1}$ and 
$X\ket{1}=\ket{0}$. The Hadamard gate is \(H\defas(X+Z)/\sqrt{2}\).
When a one-qubit gate \(G\in\{X,Y,Z,H\}\) is applied to the \(u\)-th qubit of
a \(p\)-qubit register, it acts as the identity on every other qubit. The Toffoli gate acts on three distinct qubits: if \(a\) and \(b\) are its
control qubits and \(c\) is its target qubit, then
   $\operatorname{Toffoli}_{a,b\to c}^{(p)}
    \ket{z_1\cdots z_p}
    =
    \ket{
        z_1\cdots z_{c-1}
        \bigl(z_c\oplus z_az_b\bigr)
        z_{c+1}\cdots z_p
    }$, where for \(x,y\in\{0,1\}\), 
    $x\oplus y\defas (x+y)\bmod 2$.

\subsection{Promise Problems}
A \emph{promise problem} is a pair $ \Pi=(\Pi_{\mathrm{YES}},\Pi_{\mathrm{NO}})$, where $\Pi_{\mathrm{YES}},\Pi_{\mathrm{NO}}\subseteq \bigcup_{k\in \mathbb{Z}_{+}}\{0,1\}^k$ are disjoint:
$\Pi_{\mathrm{YES}}\cap\Pi_{\mathrm{NO}}=\varnothing$. The set
    $\mathrm{dom}(\Pi)\defas
    \Pi_{\mathrm{YES}}\cup\Pi_{\mathrm{NO}}$
is called the \emph{promise}. An input in $\Pi_{\mathrm{YES}}$ is a
\emph{yes-instance}, and an input in $\Pi_{\mathrm{NO}}$ is a
\emph{no-instance}. Inputs outside $\mathrm{dom}(\Pi)$ violate the promise.
 A promise problem permits inputs that lie outside both $\Pi_{\mathrm{YES}}$ and $\Pi_{\mathrm{NO}}$. An algorithm for a promise problem is required to be correct only on inputs that satisfy the promise; its behavior on all other inputs is unrestricted \cite{watrous2009,janzing2007}.
\begin{definition}[$\BPP$]
A promise problem $\Pi$ belongs to $\BPP$ if there exists a randomized
classical algorithm $\mathcal{A}$ running in time polynomial in the
input length $|x|$ such that
$\Pr[\mathcal{A}(x)=1]\geq {2}/{3}$ for $x\in\Pi_{\mathrm{YES}}$, and $\Pr[\mathcal{A}(x)=0]\geq {2}/{3}$ for $x\in\Pi_{\mathrm{NO}}$. No condition is imposed when
$x\notin \Pi_{\mathrm{YES}}\cup\Pi_{\mathrm{NO}}$.
\end{definition}
\begin{definition}[\(\BQP\)]
\label{def:PromiseBQP}
A promise problem \(\Pi=(\Pi_{\mathrm{YES}},\Pi_{\mathrm{NO}})\), with
\(\Pi_{\mathrm{YES}}\cap\Pi_{\mathrm{NO}}=\varnothing\), belongs to
\(\BQP\) if there exist polynomials \(q,\ell:\mathbb{N}\to\mathbb{N}\),
with \(q(r)\geq r\), and a polynomial-time uniform family of quantum
circuits \(\{Y_r\}_{r\geq1}\) over a fixed finite universal gate set of
complex unitary gates, such that each \(Y_r\) acts on \(q(r)\) qubits and
has at most \(\ell(r)\) gates. On input \(x\in\{0,1\}^r\), the circuit
starts in \(\ket{x}\otimes\ket{\mathbf{0}}\), where the last \(q(r)-r\)
qubits are ancillas initialized to \(\ket{0}\), and accepts if a designated
output qubit is measured as \(1\). Writing \(p_{Y_r}(x)\) for its
acceptance probability, the circuit family satisfies
\(p_{Y_r}(x)\geq 2/3\) for every \(x\in\Pi_{\mathrm{YES}}\) and
\(p_{Y_r}(x)\leq 1/3\) for every \(x\in\Pi_{\mathrm{NO}}\). No condition is
imposed on inputs outside \(\Pi_{\mathrm{YES}}\cup\Pi_{\mathrm{NO}}\).
\end{definition}

It is immediate that $\BPP\subseteq\BQP$, since a quantum
computer can efficiently simulate any randomized classical computation.
Our conclusions about classical intractability are conditional on the
conjecture $\BPP\ne\BQP$; the completeness proofs do not assume it. As usual,
entries of the fixed elementary gate sets are computable to $k$ bits in time
polynomial in $k$.

The following lemma, proved in \cref{app:Circuits}, represents uniform polynomial-time quantum circuits by real circuits with rational entries.
\begin{lemma}[Rational approximation of quantum circuits]
\label{thm:ComplexCircuitRationalApproximation}
Let \(\mathcal{G}\) be a fixed finite gate set of constant-arity complex
unitaries, and let \(\{Y_r\}_{r\geq1}\) be a polynomial-time uniform family
over \(\mathcal{G}\), with \(q(r)\) qubits and at most \(\ell(r)\) gates,
where \(q\) and \(\ell\) are polynomially bounded. For every polynomially
bounded, integer-valued \(p\) with \(p(r)\geq1\), there exists a
polynomial-time uniform circuit family \(\{\widetilde{Y}_r\}_{r\geq1}\)
such that each \(\widetilde{Y}_r\) acts on \(q(r)+a(r)\) qubits, for some
polynomially bounded \(a\), and every gate acts on at most three qubits.
Each gate is real orthogonal, has entries in
\(\{0,\pm 1/2,\pm1\}\), and has at most four nonzero entries in each
row and column. For every \(x\in\{0,1\}^r\), the acceptance probabilities
satisfy
\(\lvert p_{\widetilde{Y}_r}(x)-p_{Y_r}(x)\rvert\leq 2/p(r)\).
\end{lemma}

\begin{definition}[Karp reduction]
Let $\Pi$ and
$\Gamma$ be promise problems.
We say that $\Pi$ is 
\emph{Karp reducible} to $\Gamma$ if there exists a deterministic classical
algorithm that computes, in time polynomial in $|x|$, a map
$f:\{0,1\}^*\to\{0,1\}^*$ such that
$x\in\Pi_{\mathrm{YES}}\Rightarrow f(x)\in\Gamma_{\mathrm{YES}}$ and
$x\in\Pi_{\mathrm{NO}}\Rightarrow f(x)\in\Gamma_{\mathrm{NO}}$.
\end{definition}

\begin{definition}[$\BQP$ hardness and completeness]
A promise problem $\Gamma$ is \emph{$\BQP$-hard} if  $\Pi$ is Karp reducible to $\Gamma$
 for every $\Pi\in\BQP$. It is \emph{$\BQP$-complete} if it is both
$\BQP$-hard and a member of $\BQP$.
\end{definition}
\subsection{Riesz Projectors and Riccati Solution}
\label{subsec:riesz-riccati}

Spectral projectors provide a way to solve Riccati equations.

Let $S\in\mathbb C^{N\times N}$, and let $\Gamma$ be a positively
oriented simple closed rectifiable contour containing no eigenvalue
of $S$. The Riesz projector associated with the eigenvalues
inside $\Gamma$ is
\begin{equation}\label{eq:Riesz_Projector}
 \Pi_\Gamma(S)
 \defas
 \frac{1}{2\pi\mathrm i}
 \oint_\Gamma(zI-S)^{-1}\,\mathrm dz.
\end{equation}
The contour-integral definition remains valid when $S$ is defective. If $S$ has no eigenvalues on the imaginary axis, a contour
enclosing precisely its left-half-plane eigenvalues defines
the stable Riesz projector, denoted by $\Pi_-(S)$.

Recall the algebraic Riccati equation and the definitions of $G$ and $Q$ in \eqref{eq:Riccati}. Assume $(A,B)$ is stabilizable and $(C,A)$ is detectable. Then, \eqref{eq:Riccati} has a unique symmetric
positive-semidefinite stabilizing solution $P_\star$; that is,
$A_\star=A-GP_\star$ is Hurwitz.

Define the Hamiltonian matrix
\begin{equation}
 \mathcal H
 \defas
 \begin{bmatrix}
  A&-G\\
  -Q&-A^\intercal
 \end{bmatrix}.
 \label{eq:lqr-hamiltonian}
\end{equation}
Let $\Pi=\Pi_-(\mathcal H)$, and partition its complete block rows as
\[
 \Pi=\begin{bmatrix}U\\W\end{bmatrix},
 \qquad U,W\in\mathbb C^{n\times2n}.
\]
Then $U$ has full row rank and
$
 P_\star=WU^+,
$
where $U^+$ denotes the Moore--Penrose inverse (see, e.g., \cite{roberts1980linear}). This representation reduces Riccati solution to three
operations: constructing the stable spectral projector, selecting
its two block rows, and applying a pseudoinverse.
\section{Problem Formulation}
\label{sec:ProblemFormulation}
Before introducing the problems of interest in this work, we highlight how the input matrices are presented. To represent high-dimensional system matrices compactly, we specify their
nonzero entries through Boolean lookup circuits rather than an explicit
entrywise listing. The following convention accommodates both square
and rectangular matrices, as required for the dynamics, input, output,
and control-weight matrices considered in this work.
\subsection{Succinct Sparse-Matrix Input Model}
\label{subsec:sparse-input}

Let $S\in\mathbb R^{d_1\times d_2}$, where $d_1,d_2\ge1$.
We call $S$ \emph{$s$-sparse} if each row and column contains
at most $s$ nonzero entries; the permitted growth of $s$
is specified in each problem definition.

For $k\in\intcc{1;2^q}$, let $\operatorname{bin}_q(k)$ denote
the $q$-bit binary representation of $k-1$.
Define
$w=\left\lceil\log_2\!\left(1+\max\{d_1,d_2\}\right)\right\rceil$,
$q_{\mathrm r}=\max\{1,\lceil\log_2 d_1\rceil\}$,
and
$q_{\mathrm c}=\max\{1,\lceil\log_2 d_2\rceil\}$.
The slot and output index widths are
$d_s=\max\{1,\lceil\log_2 s\rceil\}$,
$d_{\mathrm r}=\lceil\log_2(d_1+s)\rceil$, and
$d_{\mathrm c}=\lceil\log_2(d_2+s)\rceil$,
with the latter accommodating out-of-range padding indices.

Entry words use a fixed, efficiently decodable dyadic encoding
$\operatorname{dec}_{\nu}:\{0,1\}^{\nu}\longrightarrow\mathbb Q$,
whose values have the form $a2^{-q}$, with $a\in\mathbb Z$
and $q$ a nonnegative integer.
Fix $\eta_0\in\{0,1\}^{\nu}$ such that
$\operatorname{dec}_{\nu}(\eta_0)=0$.

The input supplies Boolean row- and column-lookup circuits:
\begin{align*}
 \mathsf{Row}_S:\;
 \{0,1\}^{q_{\mathrm r}}\times\{0,1\}^{d_s}
 &\longrightarrow
 \{0,1\}^{d_{\mathrm c}}\times\{0,1\}^{\nu},\\
 \mathsf{Col}_S:\;
 \{0,1\}^{q_{\mathrm c}}\times\{0,1\}^{d_s}
 &\longrightarrow
 \{0,1\}^{d_{\mathrm r}}\times\{0,1\}^{\nu}.
\end{align*}
Each returns the opposite index and an entry word.

If row $i\in\intcc{1;d_1}$ contains $m_i\le s$ nonzero
entries, slots $r\in\intcc{1;m_i}$ enumerate them exactly once:
\begin{align*}
 \mathsf{Row}_S
 \bigl(\operatorname{bin}_{q_{\mathrm r}}(i),
       \operatorname{bin}_{d_s}(r)\bigr)
 &=
 \bigl(\operatorname{bin}_{d_{\mathrm c}}(j),\eta\bigr),\\ S_{ij}&=\operatorname{dec}_{\nu}(\eta)\ne0.
\end{align*}
Unused slots $r\in\intcc{m_i+1;s}$ return
\[
 \mathsf{Row}_S
 \bigl(\operatorname{bin}_{q_{\mathrm r}}(i),
       \operatorname{bin}_{d_s}(r)\bigr)
 =
 \bigl(\operatorname{bin}_{d_{\mathrm c}}(d_2+r),\eta_0\bigr).
\]

Similarly, if column $j\in\intcc{1;d_2}$ contains $k_j\le s$
nonzero entries, slots $r\in\intcc{1;k_j}$ enumerate them
exactly once:
\begin{align*}
 \mathsf{Col}_S
 \bigl(\operatorname{bin}_{q_{\mathrm c}}(j),
       \operatorname{bin}_{d_s}(r)\bigr)
 &=
 \bigl(\operatorname{bin}_{d_{\mathrm r}}(i),\eta\bigr),\\
 S_{ij}&=\operatorname{dec}_{\nu}(\eta)\ne0.
\end{align*}
Unused slots $r\in\intcc{k_j+1;s}$ return
\[
 \mathsf{Col}_S
 \bigl(\operatorname{bin}_{q_{\mathrm c}}(j),
       \operatorname{bin}_{d_s}(r)\bigr)
 =
 \bigl(\operatorname{bin}_{d_{\mathrm r}}(d_1+r),\eta_0\bigr).
\]
Thus unused slots have distinct out-of-range indices within
each row or column.
Outputs on invalid row, column, or slot addresses are
unrestricted and are not queried.

The circuits must enumerate the same nonzero entries with
identical decoded values.
Both directions are supplied because efficient row access
need not imply efficient column access.
Each input matrix has its own dimensions, slot bound,
entry-word length, and circuits.

In the \emph{succinct input model}, slot bounds, entry-word
lengths, and circuit-description lengths are polynomial
in the logarithms of the matrix dimensions, with precise
bounds given in each problem definition.
For $d_1=d_2=n=2^\ell$ and $\ell\ge1$, both input address
widths equal $\ell$.
This representation follows sparse-access quantum algorithms
\cite{gilyen2019} and succinct matrix problems
\cite{papadimitriou1986,janzing2007}; its circuits can be
exponentially shorter than an explicit listing of the entries.

We now define two promise problems for scalar infinite-horizon
performance evaluation under this model: an autonomous
Lyapunov value and an optimal LQR value.

\begin{definition}[SLV]
\label{def:slv}
Fix positive integers $s,d$.
An input consists of $n=2^\ell$, with $\ell\ge1$, written in binary;
row- and column-lookup circuits for $A\in\mathbb R^{n\times n}$;
indices $i_b,i_c\in\intcc{1;n}$ specifying $b=e_{i_b}$ and $c=e_{i_c}$;
a positive rational $\gamma$; and rational thresholds $0<a_0<a_1$.
Define $\Delta=a_1-a_0$.
Each circuit description, output word, encoding of $\gamma$,
and threshold encoding has at most $\ell^d$ bits. The lookup circuits are promised to consistently describe an
$s$-sparse real matrix with dyadic entries, as specified in
Section~\ref{sec:ProblemFormulation}-\ref{subsec:sparse-input}, satisfying
\begin{equation}
 \|A\|\le1,~
 -\tfrac12(A+A^\intercal)\succeq\gamma I,~
 \max\{\gamma^{-1},\Delta^{-1}\}\le\ell^d.
 \label{eq:promises}
\end{equation}
Let $P$ be the unique solution of \eqref{eq:Lyapunov} with
$C=c^\intercal$.
Under the additional promise that one of the following alternatives
holds, the task is to decide which:
\[
 \textsc{no}:\ b^\intercal Pb\le a_0,
 \qquad
 \textsc{yes}:\ b^\intercal Pb\ge a_1.
\]
Inputs violating any representation, matrix, or value promise
lie outside the problem's promise.
\end{definition}

The dissipation condition in \eqref{eq:promises} ensures that $A$
is Hurwitz and hence that $P$ is well-defined. 

A single fixed exponent $d$ accommodates both the representation-size
and inverse-dissipation bounds, after increasing it if necessary.
For fixed $s,d$, the total encoded input length $N_{\rm in}$ satisfies
\(
 \ell\le N_{\rm in}=O(\ell^d).
\)
Consequently, a runtime polynomial in the succinct input length
is polynomial in $\log_2 n$, although an explicit listing of the
matrix entries can be exponentially larger.

We next extend scalar performance evaluation to infinite-horizon
linear-quadratic optimal control. 

\begin{definition}[$\textnormal{\textsc{SLQRV}}$]
\label{def:lqrv}
Fix an integer $D\ge1$. An input consists of
$n=2^\ell$, with $\ell\ge1$, and $p,m\ge1$, written in binary;
row- and column-lookup circuits for  $A\in\mathbb R^{n\times n}$,
 $B\in\mathbb R^{n\times p}$,
 $C\in\mathbb R^{m\times n}$, and
 $R\in\mathbb R^{p\times p}$; an index $i_b\in\intcc{1;n}$ specifying $b=e_{i_b}$;
and rational thresholds $a_0<a_1$. Define
\begin{align*}
 \Delta&=a_1-a_0,\\
 w&=\left\lceil\log_2(1+\max\{n,p,m\})\right\rceil,\\
 h&=(2+w)^D.
\end{align*}

Each matrix has a specified slot bound of at most $h$.
Each circuit description, entry word, and threshold encoding
has at most $h$ bits.

The lookup circuits are promised to consistently describe real
matrices with dyadic entries according to
Section~\ref{sec:ProblemFormulation}-\ref{subsec:sparse-input}. Furthermore,
$R=R^\intercal\succ0$ and
\begin{equation}
 \max\{\|A\|,\|B\|,\|C\|,\|R\|,\|R^{-1}\|,\Delta^{-1}\}
 \le h.
 \label{eq:lqr-data}
\end{equation}
The data also satisfy a quantitative stabilizability and
detectability promise: there exist real gains
$F\in\mathbb R^{p\times n}$ and $L\in\mathbb R^{n\times m}$
such that
\[
 \max\{\|F\|,\|L\|\}\le h
\]
and, for every $t\ge0$,
\begin{equation}
 \max\{\|e^{(A-BF)t}\|,\|e^{(A-LC)t}\|\}
 \le h e^{-t/h}.
 \label{eq:lqr-control-promise}
\end{equation}
Under the additional promise that one of the following alternatives
holds, the task is to decide which:
\[
 \textsc{no}:\ V(b)\le a_0,
 \qquad
 \textsc{yes}:\ V(b)\ge a_1,
\]
where $V$ is defined in \eqref{eq:ValueFunction}.
Any input violating the representation, matrix, or value promises
lies outside the problem's promise.
\end{definition}

The gains $F,L$ are existential witnesses, not part of the input.
Their existence ensures that the stabilizing Riccati solution
and the optimal value are well defined.
The common envelope $h$ bounds the representation sizes, sparsity,
matrix and gain norms, transient amplification, inverse decay rates,
and inverse decision gap.
No rank condition is imposed on $B$ or $C$, and $A$ need not be
Hurwitz. Neither a stabilizing policy nor an oracle for $R^{-1}$
is supplied.

The next proposition establishes that SLV is a special case
of this optimal-value problem, including its representation
and quantitative promises.

\begin{proposition}[Embedding SLV into SLQRV]
\label{prop:Embedding}
For every fixed pair of positive integers $s,d$, there exists
an integer $D_0\ge1$ such that, for every fixed $D\ge D_0$,
a classical polynomial-time map sends every promised SLV instance
with parameters $s,d$ to a promised
$\textnormal{\textsc{SLQRV}}$ instance with the same value,
thresholds, and YES or NO answer.
The map preserves $n,A,b,a_0,a_1$ and sets
\begin{equation}
 p=m=1,~
 B=0_{n\times1},~
 C=c^\intercal,~
 R=[1].
 \label{eq:slv-lqrv-embedding}
\end{equation}
\end{proposition}
\begin{proof}
Retain $n,A,b,a_0,a_1$ and apply
\eqref{eq:slv-lqrv-embedding}. Since $p=m=1$ and $n=2^\ell$, $w=\left\lceil\log_2(1+2^\ell)\right\rceil=\ell+1$ and $h=(\ell+3)^D$. Set $\Delta=a_1-a_0>0$.

We first verify the representation and data promises.
The lookup circuits for $A$ are retained, while
$B=0_{n\times1}$, $C=e_{i_c}^\intercal$, and $R=[1]$
admit consistent dyadic row and column lookups with slot bound one.
The additional circuits use only constant outputs and comparisons
with $i_c$. They can be constructed using $O(\ell)$ gates,
with descriptions of length $O(\ell\log(\ell+2))$ under a standard
explicit circuit encoding.
The retained circuit descriptions, entry words, and threshold
encodings each have at most $\ell^d$ bits by the SLV
representation promise.
Thus, there exists a constant $K_0\ge1$, depending only on
the encoding convention, such that every output circuit
description, entry word, and threshold encoding has length at most $K_0(\ell+3)^{d+1}$. Choose $D_0\ge d+1$ sufficiently large that
 $4^{D_0-(d+1)}\ge K_0$ and
 $4^{D_0}\ge\max\{s,1\}$. This choice is independent of the input thresholds.
Since $\ell+3\ge4$, for every $D\ge D_0$,
 $(\ell+3)^{D-(d+1)}
 \ge4^{D-(d+1)}
 \ge4^{D_0-(d+1)}
 \ge K_0$. Consequently,
 $K_0(\ell+3)^{d+1}
 \le(\ell+3)^{d+1}(\ell+3)^{D-(d+1)}
 =h$. Moreover,  $h=(\ell+3)^D
 \ge4^D
 \ge4^{D_0}
 \ge\max\{s,1\}$ and
 $h\ge(\ell+3)^d\ge\ell^d$. The SLV numerical promises therefore imply  $\max\{\gamma^{-1},\Delta^{-1}\}\le\ell^d\le h$. Thus, all slot and representation bounds hold.
Together with
 $R=R^\intercal\succ0$,
 $\|A\|\le1$,
 $\|B\|=0$, and
 $\|C\|=\|R\|=\|R^{-1}\|=1$, these bounds establish \eqref{eq:lqr-data}.

To verify the control promise, take
$F=0_{1\times n}$ and $L=0_{n\times1}$.
The SLV dissipation bound gives
 $\|e^{At}\|\le e^{-\gamma t}$ for $t\ge0$. Since $\gamma^{-1}\le h$, we obtain
\[
 \max\{\|e^{(A-BF)t}\|,\|e^{(A-LC)t}\|\}
 \le e^{-\gamma t}
 \le e^{-t/h}
 \le h e^{-t/h}.
\]
The zero gains satisfy the required norm bounds, establishing
\eqref{eq:lqr-control-promise}. Finally, because $B=0$, every admissible input yields the same
trajectory $x(t)=e^{At}b$. Consequently,
\[
 \begin{aligned}
 V(b)
 &=\inf_{u(\cdot)}
 \left[
 \int_0^\infty |c^\intercal e^{At}b|^2\,\mathrm{d}t
 +\int_0^\infty |u(t)|^2\,\mathrm{d}t
 \right]\\
 &=\int_0^\infty |c^\intercal e^{At}b|^2\,\mathrm{d}t
 =b^\intercal Pb,
 \end{aligned}
\]
with the infimum attained by $u\equiv0$.
The value and thresholds are unchanged, so the value promise
and YES or NO answer are preserved.

The map copies the original circuits and threshold encodings
and constructs additional circuits of size polynomial in $\ell$.
Since the binary encoding of $n=2^\ell$ has length $\ell+1$,
the construction runs in classical time polynomial in the
original input length.
\end{proof}
\section{Hardness Result}
\label{sec:Hardness}
In this section, we establish the hardness of SLV.
\begin{theorem}[Hardness of the \textnormal{\textsc{SLV}} problem]
\label{thm:main}
Fix $s=d=5$.
For every $\Pi\in\BQP$, there exist a polynomially bounded
integer function $L(r)$ and a polynomial-time classical map
such that, for every promised input $x\in\{0,1\}^r$, the map
outputs an \textsc{SLV} instance
\[
 \left(n(r),A_r,i_b(x),i_c(x),\gamma_r,
       a_0=\frac4{25},a_1=\frac9{25}\right).
\]
The matrix $A_r$ and rational $\gamma_r>0$ depend only on
$r$ and $\Pi$, while the coordinate indices may depend on $x$.
Writing $b_x=e_{i_b(x)}$ and $c_x=e_{i_c(x)}$, the map
preserves the promise: $x\in\Pi_{\mathrm{YES}}$ implies the output is in $\mathrm{SLV}_{\mathrm{YES}}$, and $x\in\Pi_{\mathrm{NO}}$ implies the output is in $\mathrm{SLV}_{\mathrm{NO}}$.
The reduction can be chosen so that
 \begin{align*}
 \ell(r)&=L(r)+\lceil\log_2(L(r)+1)\rceil,\\
 n(r)&=2^{\ell(r)},\\
 \alpha_r&=\Theta(\ell(r)^{-1/2}).
\end{align*}
Its outputs satisfy:
\begin{enumerate}
 \item
 $A_r\in\mathbb R^{n(r)\times n(r)}$ is $5$-sparse
 with dyadic entries, every eigenvalue equals $-\alpha_r$,
 and
  $\|A_r\|\le1$,
  $-(A_r+A_r^\intercal)/2\succeq\gamma_r I$, and
  $\max\{\gamma_r^{-1},(a_1-a_0)^{-1}\}
  \le\ell(r)^d$.

 \item
 $A_r$ has consistent row- and column-lookup circuits
 whose descriptions have at most $\ell(r)^d$ bits.
 Each output word encoding an entry of $A_r$, and each
 numerator--denominator encoding of $\gamma_r,a_0,a_1$,
 has at most $\ell(r)^d$ bits.
\end{enumerate}
Consequently, \textsc{SLV} is $\BQP$-hard even when the input
thresholds are restricted to $a_0=4/25$ and $a_1=9/25$.
\end{theorem}

\subsection{Proof of Theorem~\ref{thm:main}}
\label{sec:Proof}  
\subsubsection{Reduction From a Quantum Decision Problem}

Let \(\Pi=(\Pi_{\mathrm{YES}},\Pi_{\mathrm{NO}})\) be an arbitrary problem in
\(\BQP\), and let \(x\in\{0,1\}^r\) be an instance of \(\Pi\). Let
\(\{Y_r\}_{r\geq 1}\) be the polynomial-time uniform circuit family from
\cref{def:PromiseBQP}, whose acceptance probability \(p_{Y_r}(x)\) satisfies
$x\in\Pi_{\mathrm{YES}}\Rightarrow p_{Y_r}(x)\geq {2}/{3}$
and
$x\in\Pi_{\mathrm{NO}}\Rightarrow p_{Y_r}(x)\leq{1}/{3}$. Apply \cref{thm:ComplexCircuitRationalApproximation} with the constant
precision parameter \(p(r)=30\). This gives a polynomial-time uniform
family \(\{\widetilde{Y}_r\}_{r\geq 1}\) whose gates are real orthogonal
matrices with entries in \(\{0,\pm {1}/{2},\pm1\}\), and whose
acceptance probability differs from that of \(Y_r\) by at most \(1/15\).
Consequently, for the acceptance probability of
\(\widetilde{Y}_r\) on input \(x\), we have
    $x\in\Pi_{\mathrm{YES}}\Rightarrow p_{\tilde{Y}_{r}}(x)\geq  {3}/{5}$ and $x\in\Pi_{\mathrm{NO}}\Rightarrow p_{\tilde{Y}_{r}}(x)\leq {2}/{5}$. For notational simplicity, we henceforth write \(\widetilde{Y}_r\) as \(Y_r\).

Let \(Q(r)\) and \(T(r)\) denote, respectively, the number of qubits and
gates in \({Y}_r\); both are polynomially bounded in \(r\).
Write \({Y}_r=G_{r,T(r)}\cdots G_{r,1}\), where
\(G_{r,t}\) is the full-register matrix of the \(t\)-th gate, with
\(G_{r,1}\) applied first. Order the qubits so that the designated output
qubit is first, and define
\(\Pi_{\rm acc}\defas\ket{1}\!\bra{1}\otimes I_{2^{Q(r)-1}}\).
Let \(\ket{\widetilde{x}}\in\mathbb{R}^{2^{Q(r)}}\) denote the
computational-basis state encoding \(x\), with all other qubits initialized
to \(\ket{0}\). Then
\(p_x\defas p_{{Y}_{r}}(x)=\|\Pi_{\rm acc}{Y}_r\ket{\widetilde{x}}\|^2\).
Set \(N_r\defas 2^{Q(r)}\).  
Introduce one additional ancillary qubit \(a\). The enlarged register
consists of \(Q(r)+1\) qubits and has state-space dimension \(2N_r\). Define
\begin{equation}
    D_r
    \defas
    ({Y}_r^{\intercal}\otimes I_2)\,
    \operatorname{CNOT}_{\rm out\to a}\,
    ({Y}_r\otimes I_2)
    \in\mathbb{R}^{2N_r\times2N_r},
    \label{eq:gadget}
\end{equation}
where 
\begin{equation}  \operatorname{CNOT}_{\rm out\to a}
    \defas
    \bigl(I_{N_r}-\Pi_{\rm acc}\bigr)\otimes I_2
    +
    \Pi_{\rm acc}\otimes X.
    \label{eq:CNOT-definition}
\end{equation}
 Since \({Y}_r\) and $\operatorname{CNOT}_{\rm out\to a}$
are real orthogonal, the matrix \(D_r\) in \eqref{eq:gadget} is also real
orthogonal. Set
$\ket{\psi_{\rm in,x}}
    \defas
    \ket{\widetilde x}\otimes\ket{0}$ and
    $\ket{\psi_{\rm out,x}}
    \defas
    \ket{\widetilde x}\otimes\ket{1}$. Both states are computational-basis states. Additionally, define
$ \ket{v_x}
    \defas
    {Y}_r\ket{\widetilde x}$,
     $\ket{v_{x,0}}
    \defas
    (I_{N_r}-\Pi_{\rm acc})\ket{v_x}$, and
    $\ket{v_{x,1}}
    \defas
    \Pi_{\rm acc}\ket{v_x}$. Thus, \(\ket{v_{x,0}}\) and \(\ket{v_{x,1}}\) are respectively the
components of \(\ket{v_x}\) for which the output qubit has value \(0\) and
\(1\). By \eqref{eq:CNOT-definition},
\begin{align*}
\operatorname{CNOT}_{\rm out\to a}
    \bigl(\ket{v_x}\otimes\ket{0}\bigr)
    &=
  \ket{v_{x,0}}\otimes\ket{0}
    +
    \ket{v_{x,1}}\otimes X\ket{0}\\
    &=
\ket{v_{x,0}}\otimes\ket{0}
    +
    \ket{v_{x,1}}\otimes\ket{1}.
    \end{align*}
 Using \({Y}_r^{\intercal}={Y}_r^{-1}\), we obtain
$ \bra{\psi_{\rm out,x}}D_r\ket{\psi_{\rm in,x}}=
    p_x$.

\subsubsection{Construction of the Continuous-Time System}
The matrix \(D_r\) has a circuit decomposition into constant-size real
orthogonal gates. Recall the decomposition 
\({Y}_r= G_{r,T(r)}\cdots G_{r,1}\), and note that
\({Y}_r^{\intercal}\) is implemented by the reversed sequence of
transposed gates. Together with the controlled-NOT gate in
\eqref{eq:gadget}, this gives a decomposition of \(D_r\) with
\(L_0(r)\defas 2T(r)+1\) components, acting on
\(r_0(r)\defas Q(r)+1\) qubits, including the ancillary qubit in
\eqref{eq:gadget}.

We pad the decomposition so that its number of components is at least its
number of qubits. Set
\(L(r)\defas\max\{L_0(r),r_0(r),L_{\min}\}\), where the fixed
constant $L_{\min}$ absorbs the finite encoding overhead in
Appendix~\ref{sec:ProofContinuation}. If
\(L_0(r)<L(r)\), append \(L(r)-L_0(r)\) identity gates, yielding
\(D_r=G_{r,L(r)}\cdots G_{r,1}\) without changing the implemented matrix.
Then append \(L(r)-r_0(r)\) qubits, initialized to \(\ket{0}\), and extend
each gate by the identity on these qubits. The resulting circuit implements
\(\widehat D_r\defas D_r\otimes I_{2^{L(r)-r_0(r)}}\) on \(L(r)\) qubits.
For notational simplicity, we henceforth denote \(\widehat D_r\) by \(D_r\).
Thus \(D_r=G_{r,L(r)}\cdots G_{r,1}\), where each of its \(L(r)\) real
orthogonal components acts nontrivially on at most a constant number of
qubits and as the identity on all other qubits.

The padded circuit, and every matrix introduced below before choosing
\(b\) and \(c\), depends only on \(r\). For a fixed input \(x\), define
\(\ket{\widehat\psi_{\rm in}}\defas
\ket{\psi_{\rm in,x}}\otimes\ket{0}^{\otimes(L(r)-r_0(r))}\) and
\(\ket{\widehat\psi_{\rm out}}\defas
\ket{\psi_{\rm out,x}}\otimes\ket{0}^{\otimes(L(r)-r_0(r))}\).
Since the added qubits remain unchanged, their transition amplitude is
\(\bra{\widehat\psi_{\rm out}}D_r\ket{\widehat\psi_{\rm in}}
=\bra{\psi_{\rm out,x}}D_r\ket{\psi_{\rm in,x}}=p_x\).

For the remainder of the construction, fix \(x\) and write
\(p\defas p_x\), \(\ket{\psi_{\rm in}}\defas\ket{\widehat\psi_{\rm in}}\),
and \(\ket{\psi_{\rm out}}\defas\ket{\widehat\psi_{\rm out}}\). For
brevity, also write \(L\defas L(r)\).
    
Let \(\mathcal{H}_{\rm clk}\defas\mathbb{R}^{L+1}\) be the clock space,
with computational basis
    $\ket{0}_{\rm clk},\ldots,\ket{L}_{\rm clk}$,
and let \(\mathcal{H}_{\rm work}\defas\mathbb{R}^{2^L}\) be the state space
of the \(L\)-qubit work register.  Define the clocked propagation operator $W_{r}\in
    \mathbb{R}^{(L+1)2^L\times(L+1)2^L}$ as: \begin{equation}
    W_r
    \defas
    \sum_{k=0}^{L-1}
    \ket{k+1}_{\rm clk}\!\bra{k}_{\rm clk}
    \otimes G_{r,k+1}.
    \label{eq:clock}
\end{equation}
This is the forward, noncyclic part of the Feynman clock
construction \cite{feynman1985,kitaev2002}. 
  Each \(G_{r,\ell}\) acts nontrivially on at most a constant number of
qubits. Therefore, there exists a constant \(s\), independent of \(r\) and
\(L\), such that every row and every column of each \(G_{r,\ell}\) has at
most \(s\) nonzero entries. It follows from \eqref{eq:clock} that every row
and every column of \(W_r\) has at most \(s\) nonzero entries. Hence,
\(W_r\) is constant sparse.

For \(k\in \intcc{1; L}\), $$W_r^k
    =
    \sum_{j=0}^{L-k}
    \ket{j+k}_{\rm clk}\bra{j}_{\rm clk}
    \otimes
    G_{r,j+k}\cdots G_{r,j+1}.
    $$
    In particular, the formula for \(k=L\) reduces to
    $W_r^{L}
    =
    \ket{L}_{\rm clk}\bra{0}_{\rm clk}
    \otimes
    G_{r,L}\cdots G_{r,1}.
    $
    Therefore,
    $W_r^{L+1}
    =0$ and
 \(W_r\) is nilpotent, with nilpotency index at most \(L+1\). Moreover, \(W_r\) is a partial isometry. Indeed, because each
\(G_{r,\ell}\) is real orthogonal,
\(
    W_r^{\intercal} W_r
    =
    \sum_{k=0}^{L-1}
    \ket{k}_{\rm clk}\!\bra{k}_{\rm clk}
    \otimes I_{2^L},
\)
and, consequently,  $\|W_r\|=1$. 

Define
\begin{equation}
    \ket{b}
    \defas
    \ket{0}_{\rm clk}\otimes\ket{\psi_{\rm in}},
    \qquad
    \ket{c}
    \defas
    \ket{L}_{\rm clk}\otimes\ket{\psi_{\rm out}}.
    \label{eq:bc}
\end{equation}
For \(k<L\),
    $\bra{c}W_r^k\ket{b}=0$
. On the other hand,
$
    \bra{c}W_r^L\ket{b}
    =
    \bra{\psi_{\rm out}}
    G_{r,L}\cdots G_{r,1}
    \ket{\psi_{\rm in}} 
=p.
$
Define
\begin{align}
    \alpha_L
    &\defas
    \frac{(2L)!}{2^{2L+1}(L!)^2}
    =
    \frac{\binom{2L}{L}}{2^{2L+1}}=\Theta(L^{-1/2}),
    \label{eq:alphaL}\\
    A_r^{\rm clk}
    &\defas
    \alpha_L\bigl(-I_{(L+1)2^L}+W_r\bigr)
    \in
    \mathbb{R}^{(L+1)2^L\times(L+1)2^L}.
    \label{eq:Aclock}
\end{align}

We next verify that \(A_r^{\rm clk}\) is strictly dissipative.  The symmetric part of the
clocked propagation operator has the eigenvalues:
$$
\cos\left(\frac{j\pi}{L+2}\right),
     ~   j\in \intcc{1;L+1},
     $$ 
where each eigenvalue occurs with multiplicity \(2^L\); see
Appendix~\ref{app:clock-spectrum} for a derivation. In particular, the maximum eigenvalue of  $(W_r+W_r^\intercal)/{2}$ is
    $\cos({\pi}/(L+2))$. Therefore,
\begin{align*}
    -\frac{A_r^{\rm clk}+(A_r^{\rm clk})^\intercal}{2}
    =&
    \alpha_L
    \left(
        I-\frac{W_r+W_r^\intercal}{2}
    \right)\\
    &\succeq
    \alpha_L
    \left(
        1-\cos\left(\frac{\pi}{L+2}\right)
    \right)I.
\end{align*}
Thus, \(A_r^{\rm clk}\) is \(\gamma_L\)-dissipative, where
\begin{align}
    \gamma_L\defas
    \alpha_L
    \left(
        1-\cos\!\left(\frac{\pi}{L+2}\right)
    \right)=
    \Theta(L^{-5/2}).
    \label{eq:gammaL}
\end{align}
The asymptotic estimate for \(\gamma_L\) follows from
\(\alpha_L=\Theta(L^{-1/2})\) and
\(
    1-\cos({\pi}/{(L+2)})
    =
    \Theta(L^{-2}).
\)
In fact, the nilpotency of $W_r$ implies that $-\alpha_L$ is the only
eigenvalue of $A_r^{\rm clk}$.

To describe the associated continuous-time system, write
   $$ \ket{z(\tau)}
    =
    \sum_{k=0}^{L}
    \ket{k}_{\rm clk}\otimes\ket{z_k(\tau)},
$$
where
$
\ket{z_k(\tau)}\in\mathcal{H}_{\rm work},
$
and \(\tau\geq0\) denotes continuous physical time. Then the system  
    $$\ket{\dot{z}(\tau)}
    =
    A_r^{\rm clk}\ket{z(\tau)}
    $$
has the feed-forward form:
\begin{align*}
\ket{\dot{z}_0(\tau)}
    &=
    -\alpha_L\ket{z_0(\tau)},\\  
    \ket{\dot{z}_k(\tau)}&=
    -\alpha_L\ket{z_k(\tau)}
    +
    \alpha_LG_{r,k}\ket{z_{k-1}(\tau)},~k\in\intcc{1;L}.
\end{align*}
 By the nilpotency of $W_{r}$,
$$
\exp(A_r^{\rm clk}\tau)=\exp(-\alpha_L\tau)
\displaystyle\sum_{k=0}^L
\frac{(\alpha_L\tau)^k}{k!}W_r^k.
$$
We therefore obtain
    $$
    c^{\intercal}\exp{(A_r^{\rm clk}\tau)}b
    =
    \exp{(-\alpha_L\tau)}
    \frac{(\alpha_L\tau)^L p}{L!},
$$
    and consequently,
    \begin{align*}
    J
    &=
    \int_0^\infty
    \bigl|c^{\intercal}e^{A_r^{\rm clk}\tau}b\bigr|^2\,d\tau=
 \frac{p^2\alpha_L^{2L}}{(L!)^2}
    \int_0^\infty e^{-2\alpha_L\tau}\tau^{2L}\,d\tau \notag\\
    &=
    p^2
    \frac{(2L)!}{2^{2L+1}\alpha_L(L!)^2}=
    p^2,
\end{align*}
where we used $\int_0^\infty e^{-\beta\tau}\tau^m\,d\tau = {m!}/(\beta^{m+1})$ for $\beta>0$ and the definition of $\alpha_{L}$ in \eqref{eq:alphaL}.
Hence, a no instance satisfies \(J\leq 4/25\), whereas a yes instance
satisfies \(J\geq 9/25\).  Both ranges lie within the respective
threshold intervals in Definition~\ref{def:slv}.

\subsubsection{A Rational Dissipativity Certificate}
Although the exact dissipativity margin \(\gamma_L\) in
\eqref{eq:gammaL} is generally irrational, the SLV instance need only
contain a rational number that certifies a valid lower bound.  Since
\(\sin y\geq 2y/\pi\) for \(0\leq y\leq\pi/2\), we have
$$  1-\cos\left(\frac{\pi}{L+2}\right)
    =
    2\sin^2\!\left(\frac{\pi}{2L+4}\right)\geq \frac{2}{(L+2)^2}.
$$
    Therefore, the rational number
\begin{equation}
    \gamma_{L,\mathrm{r}}
    \defas
    \frac{2\alpha_L}{(L+2)^2}
    \leq
    \gamma_L
    \label{eq:certificate}
\end{equation}
is a valid dissipativity certificate.  Moreover, the central-binomial bound
\(
    \alpha_L\geq{1}/{(4\sqrt{L})}
\)
implies that
\(
    \gamma_{L,\mathrm{r}}^{-1}
    \leq
    2\sqrt{L}\,(L+2)^2
    =
    O(L^{5/2}).
\)

\subsubsection{Exact Encoding of the Numerical Parameters}
We next record the exact bit lengths of the numerical quantities used in
the reduction.  Recall that
\(
    \alpha_L
    =
    {\binom{2L}{L}}/{2^{2L+1}}.
\)
Because \(\binom{2L}{L}\leq 2^{2L}\), its binary representation has at most \(2L+1\)
bits.  Thus \(\alpha_L\) has an exact dyadic encoding of \(O(L)\) bits.
Every component matrix \(G_{r,\ell}\) has entries in
\(  \left\{0,\pm 1/2,\pm 1\right\}.
\)
Consequently, every entry of
\(
    A_r^{\rm clk}=\alpha_L(-I+W_r)
\)
is either zero, \(-\binom{2L}{L}/2^{2L+1}\), or one of $\pm{\binom{2L}{L}}/{2^{2L+1}}$ and
    $\pm{\binom{2L}{L}}/{2^{2L+2}}$.
Hence every entry of \(A_r^{\rm clk}\) is an exact dyadic rational whose numerator
and dyadic denominator exponent require \(O(L)\) bits. Recall the rational certificate
\(
    \gamma_{L,\mathrm{r}}
    =
    {\binom{2L}{L}}/(2^{2L}(L+2)^2).
\)
This is an exact numerator--denominator representation: its numerator has
at most \(2L+1\) bits, while its denominator has
\(
    2L+O(\log(L+2))
\)
bits.  Therefore the rational encoding of
\(\gamma_{L,\mathrm{r}}\) also has \(O(L)\) bits.  The integer \(\binom{2L}{L}\), the
dyadic encoding of \(\alpha_L\), and the numerator and denominator above
can all be computed in time polynomial in \(L\) using ordinary integer
arithmetic.

\subsubsection{Sparsity and Norm Bounds for the Clocked Matrix}
Each row and each column of \(W_r\) contains the entries of at most one
component matrix \(G_{r,\ell}\). Since every row and column of
\(G_{r,\ell}\) has at most four nonzero entries, every row and column of
\(A_r^{\rm clk}\) has at most five nonzero entries: its diagonal entry and
at most four off-diagonal entries inherited from \(W_r\).  Thus,
\(A_r^{\rm clk}\) is \(5\)-sparse.
Finally, since \(\|W_r\|=1\) and \(\alpha_L\leq1/4\),
\(
    \|A_r^{\rm clk}\|
    \leq
    \alpha_L\bigl(\|I\|+\|W_r\|\bigr)
    =
    2\alpha_L
    \leq
    1/2.
\)

\subsubsection{Padding to a Power-of-Two Dimension}
Recall that the \textsc{SLV} input model encodes row and column indices by
binary strings of a common length. We therefore pad the clocked construction
to a power-of-two dimension, so that every state coordinate is indexed by an
\(\ell\)-bit string. Set
   $ m\defas (L+1)2^L$, 
    $\ell
    \defas
    L+\left\lceil\log_2(L+1)\right\rceil$,
    and
   $ n\defas2^\ell$.
Thus, \(m\leq n\).  Define
    $\overline W_r
    \defas
    W_r\oplus0_{n-m}$,
and
   $ \overline A_r
    \defas
    \alpha_L\bigl(-I_n+\overline W_r\bigr) \notag=
    A_r^{\rm clk}\oplus\bigl(-\alpha_L I_{n-m}\bigr)$. Extend \(\ket{b}\) and \(\ket{c}\) by zero coordinates, namely,
    $\ket{\tilde b}
    \defas
    \begin{bmatrix}
        \bra{b},
        0_{n-m}^{\intercal}
    \end{bmatrix}^{\intercal}$ and 
    $\ket{\tilde c}
    \defas
    \begin{bmatrix}
        \bra{c},
        0_{n-m}^{\intercal}
    \end{bmatrix}^{\intercal}$
and, for notational simplicity, continue to denote these extended vectors by
\(\ket{b}\) and \(\ket{c}\). The added coordinates are dynamically decoupled from the original clocked
system. Indeed, if $\ket{z(\tau)}=\bigl(\bra{z_{\rm clk}(\tau)},\,\bra{z_{\rm pad}(\tau)}\bigr)^{\intercal}$,
then the system
\(
   \ket{\dot{z}(\tau)}
    =
    \overline A_r\ket{z(\tau)}
\)
satisfies
    $\ket{\dot{z}_{\rm clk}(\tau)}
    =
    A_r^{\rm clk}\ket{z_{\rm clk}(\tau)}$,
    $\ket{\dot{z}_{\rm pad}(\tau)}
    =
    -\alpha_L\ket{z_{\rm pad}(\tau)}$.
Thus, the added coordinates remain zero when initialized at zero and do not
affect the original clocked dynamics. In particular, $\bra{\tilde{c}}\overline W_r^L\ket{\tilde{b}}
    =
    \bra{c}W_r^L\ket{b}
    =
    p$. The padding also preserves the nilpotence of the clocked propagation
operator and the norm bound.  It also preserves the dissipativity
certificate: on the padded subspace the negative symmetric part is
\(\alpha_L I\), and \(\alpha_L\geq\gamma_{L,\mathrm{r}}\).  Thus
\(-(\overline A_r+\overline A_r^\intercal)/{2}
    \succeq
    \gamma_{L,\mathrm{r}} I.
\)
Moreover, the padding adds only the eigenvalue \(-\alpha_L\) and diagonal
entries \(-\alpha_L\).  Hence the sparsity, dyadic-entry, bit-length, and
eigenvalue properties already proved for the clocked block hold for
\(\overline A_r\) as well.

The remaining step constructs polynomial-size row- and column-lookup circuits for \(\overline A_r\), and verifies that the reduction satisfies the sparsity, encoding-length, and dissipativity promises of \textsc{SLV}. This involves tedious bookkeeping of clock and work addresses, gate entries, and padding; the detailed construction and verification continue in \cref{sec:ProofContinuation}.

\section{Quantum Membership for the Infinite-Horizon LQR Value}
\label{sec:Membership}
The previous section establishes $\BQP$-hardness through SLV, a special case of SLQRV. To establish $\BQP$-completeness, we now analyze the query complexity of the general SLQRV problem and show that it belongs to $\BQP$.  

\begin{theorem}[Membership for SLQRV]
\label{thm:lqrv-membership}
For every fixed integer $D\ge1$, $\textsc{SLQRV}$
belongs to $\BQP$.
More precisely, given rational accuracy parameters
$0<\varepsilon\le1$ and $0<\delta<1/2$, there is
a uniform quantum algorithm that estimates the
optimal value $V(b)$ to additive error $\varepsilon$,
with probability at least $1-\delta$, in time
polynomial in the total encoded input length,
$1/\varepsilon$, and $\log(1/\delta)$.
The total input length includes the encodings
of the accuracy parameters.
\end{theorem}

The proof implements the quantum CARE construction
of \cite[Theorem~3.5]{wang2026near} from the supplied
succinct matrix descriptions.
The following lemma supplies the solution-norm and
resolvent bounds required by that construction.
Its proof appears in
Appendix~\ref{sec:Proof_lem:lqr-conditioning}.

\begin{lemma}[Quantitative conditioning bounds]
\label{lem:lqr-conditioning}
Consider an instance satisfying the representation,
data, and control promises of \cref{def:lqrv}.
Let $P_\star$ be the unique positive semidefinite
stabilizing solution of \eqref{eq:Riccati}, and
let $\mathcal H$ be the Hamiltonian in
\eqref{eq:lqr-hamiltonian}.
Then
\begin{align}
 \|P_\star\|
 &\le p_0\defas h^6,
 \label{eq:lqr-P-bound}\\
 \sup_{\omega\in\mathbb R}
 \|(i\omega I-\mathcal H)^{-1}\|
 &\le M\defas 2h^2+4h^7.
 \label{eq:lqr-axis-bound}
\end{align}
In particular, $\mathcal H$ has no eigenvalues
on the imaginary axis.
\end{lemma}

\subsection{Proof of
Theorem~\ref{thm:lqrv-membership}}

We construct block-encoding access to the Hamiltonian,
apply the quantum CARE construction, and estimate
$V(b)=b^\intercal P_\star b$.

A block encoding of a matrix $S$ with normalization
$\alpha_S>0$ and error $\eta$ is a unitary $U_S$
with specified input and output isometries
$J_{\mathrm{in}},J_{\mathrm{out}}$ such that
 $\widehat S
 :=\alpha_S
 J_{\mathrm{out}}^\dagger U_SJ_{\mathrm{in}}$ satisfies
 $\|\widehat S-S\|\le\eta$. Thus $\widehat S$ is the matrix represented by
the encoding, and every error tolerance below
refers to its distance from the target matrix.
For rectangular $S$, the input and output
isometries may have different domain dimensions.

\subsubsection{Sparse Matrix Access}
The supplied lookup circuits specify the
locations and exact dyadic values of the
nonzero entries of $A,B,C,R$.
Their reversible implementations provide
coherent access with polynomial gate overhead.
When access to the entry at a specified pair
of indices $(i,j)$ is required, we scan the
at most $h$ slots in row $i$, returning the
matching value or zero.
Such an entry query therefore requires
$O(h)$ evaluations of the supplied lookup
circuits, including uncomputation.

For each $S\in\{A,B,C,R\}$, every row and
column contains at most $h$ nonzero entries,
each of absolute value at most $h$, since $|S_{ij}|\le\|S\|\le h$.
Consequently,
 $\|S\|
 \le\sqrt{\|S\|_1\|S\|_\infty}
 \le h^2$. Separately, the promise $\|R^{-1}\|\le h$ gives $R\succeq h^{-1}I$.

The sparse block-encoding construction
\cite[Lemma~48]{gilyen2019}
permits normalization $h^2$ under these
sparsity and entry bounds.
We use the common normalization
$\alpha_A=\alpha_B=\alpha_C=\alpha_R
 =:\alpha=2h^2$,
where the constant factor provides convenient
slack and is independent of the target accuracy.
Each sparse block-encoding call requires
$O(h)$ lookup-circuit evaluations.

\subsubsection{Padding}
If padding is needed, extend $B$ and $C$ by
zeros and $R$ by an identity.
Extend $A$ by $-I$ on any additional state
coordinates.
The added state coordinates are decoupled,
and the stabilizing Riccati solution extends
as $P_\star\oplus0$.
Their Hamiltonian blocks have eigenvalues
$\pm1$ and imaginary-axis resolvent norm
at most one.
Thus the bounds in
Lemma~\ref{lem:lqr-conditioning} remain valid.
We suppress these extensions in the notation.

\subsubsection{Hamiltonian Construction and Error Budget}
Our goal is the following: for any prescribed
$0<\xi\le1$, construct a block encoding of
$\mathcal H$ with normalization
$\alpha_{\mathcal H}=O(h^5)$ and error at
most $\xi$.
We allocate half of this error budget to
approximating $R^{-1}$ and half to implementing
the composed circuit.

Since $R\succeq h^{-1}I$, the smallest singular
value of $R/\alpha$ is at least
$1/(\alpha h)$.
Inverse QSVT
\cite[Theorem~41]{gilyen2019}
therefore constructs an ideal encoding of
a matrix $\widehat S_R$ satisfying
$\|\widehat S_R-R^{-1}\|\le \rho$ with known normalization  $\beta=c_Rh=O(h)$,
where $c_R$ is a sufficiently large absolute
constant.
The number of calls to the ideal encoding
of $R$ and its adjoint is
\[
 O\!\left(
 \alpha h\log\frac{2h}{\rho}
 \right)
 =
 O\!\left(
 h^3\log\frac{2h}{\rho}
 \right).
\]
The factor $\alpha h=O(h^3)$ is the reciprocal
of the supplied lower bound on the smallest
singular value of the normalized input.

Using adjoint, product, and linear-combination
constructions
\cite[Lemmas~52--53]{gilyen2019},
we obtain ideal encodings of $ \mathcal H_\rho$, defined as in \eqref{eq:lqr-hamiltonian} by replacing $G$ with
 $G_\rho:=B\widehat S_RB^\intercal$, with known normalizations
 $\alpha_G=\alpha^2\beta=O(h^5)$,
 $\alpha_Q=\alpha^2=O(h^4)$, and
 $\alpha_{\mathcal H}=2\alpha+\alpha_G+\alpha_Q=O(h^5)$. These normalizations are fixed independently
of $\xi$.
Neither $R^{-1}$ nor $G$ needs to be sparse,
because their encodings are obtained by
inverse and product constructions.

The only matrix approximation at this ideal
stage is the replacement of $R^{-1}$ by
$\widehat S_R$.
Since $\|B\|\le h^2$,
$
 \|\mathcal H_\rho-\mathcal H\|
 =\|B(\widehat S_R-R^{-1})B^\intercal\|\le\|B\|^2\rho
 \le h^4\rho$. We therefore choose
\[
 \rho:=\frac{\xi}{2h^4},
\]
which ensures
$\|\mathcal H_\rho-\mathcal H\|\le\xi/2$.

Let $U_{\mathcal H,\rho}$ denote the ideal
composed encoding of $\mathcal H_\rho$.
Implement its arithmetic, entry-dependent
rotations, and QSVT phases to obtain
$\widehat U_{\mathcal H}$ satisfying
 $\|\widehat U_{\mathcal H}
       -U_{\mathcal H,\rho}\|
 \le {\xi}/(2\alpha_{\mathcal H})$.

Such precision is achievable with polynomial
overhead.
Indeed, for a circuit containing
$L_{\mathrm{circ}}$ elementary operations,
an error at most
$\xi/(2\alpha_{\mathcal H}L_{\mathrm{circ}})$
per operation suffices by the telescoping
bound for products of unitaries.
The required precision therefore grows only
logarithmically with $L_{\mathrm{circ}}$,
$\alpha_{\mathcal H}$, and $1/\xi$.

Define the matrix represented by the
implemented encoding as
\[
 \widehat{\mathcal H}
 :=
 \alpha_{\mathcal H}
 J_{\mathrm{out}}^\dagger
 \widehat U_{\mathcal H}
 J_{\mathrm{in}}.
\]
Compression by isometries cannot increase
the operator norm, so
 $\|\widehat{\mathcal H}-\mathcal H_\rho\|
 \le
 \alpha_{\mathcal H}
 \|\widehat U_{\mathcal H}
       -U_{\mathcal H,\rho}\|
 \le {\xi}/{2}$. Combining the inverse-approximation and
implementation errors gives
\(
 \|\widehat{\mathcal H}-\mathcal H\|
 \le\xi.
\)

\subsubsection{Cost of Hamiltonian Access}
Let $Q_{\mathcal H}(\xi)$ denote the maximum
number of evaluations of the supplied lookup
circuits, including their reversible inverses,
required for one call to
$\widehat U_{\mathcal H}$, its adjoint,
or their controlled versions.

The inverse construction uses
$O(h^3\log(2h/\rho))$ sparse block-encoding calls,
each costing $O(h)$ lookup-circuit evaluations.
The remaining product and linear-combination
constructions add only a constant number of
sparse block-encoding calls.
Substituting $\rho=\xi/(2h^4)$ therefore gives
\[
 Q_{\mathcal H}(\xi)
 =O\!\left(
 h^4\log\frac{2h}{\xi}
 \right).
\]
Thus, for every $0<\xi\le1$, we obtain a
Hamiltonian block encoding with known
normalization $\alpha_{\mathcal H}=O(h^5)$,
represented-matrix error at most $\xi$,
and the displayed lookup-query cost per call.
The gate cost per call is polynomial in
the encoded instance length, $h$, and
$\log(1/\xi)$.

\subsubsection{Stable-Projector Construction}
We next construct a block encoding of the stable
Riesz projector of $\mathcal H$ and use it to recover $P_\star$.

By Lemma~\ref{lem:lqr-conditioning}, we have
the known bounds
 $\|P_\star\|\le p_0:=h^6$ and
 $\sup_{\omega\in\mathbb R}
 \|(i\omega I-\mathcal H)^{-1}\|\le M\defas  2h^2+4h^7.
\)
The second bound supplies the imaginary-axis
singular-value lower bound
$\underline\Delta_{\mathrm{ax}}=M^{-1}$
required by
\cite[Proposition~F.3]{wang2026near}.
Let 
\begin{align*}
 \sigma&:=\frac1{2M},\\
\Gamma_- &\defas \{x+iy:
   -2\alpha_{\mathcal H}\le x\le-\sigma,\quad
   |y|\le2\alpha_{\mathcal H}\}.
\end{align*}
The cited proposition guarantees that this
contour encloses exactly the eigenvalues of
$\mathcal H$ with negative real part and that
\begin{align*}
 &s_\Gamma
 :=\operatorname{length}(\Gamma_-)
 =12\alpha_{\mathcal H}-2\sigma
 \le12\alpha_{\mathcal H},\\
 &\sup_{z\in\Gamma_-}
 \|(zI-\mathcal H)^{-1}\|\le 2M.
\end{align*}
Thus, the  Riesz projector $\Pi$ defined over $\Gamma_{-}$ as in \eqref{eq:Riesz_Projector} can be used to recover $P_{\star}$. 
We approximate $\Pi$  using the
panelwise Gauss--Legendre rule of
\cite[Proposition~F.3]{wang2026near}.
For quadrature nodes $z_1,\ldots,z_m$ on
$\Gamma_-$ and complex coefficients
$\varpi_1,\ldots,\varpi_m$, write
\[
 S_m:=\sum_{j=1}^m
       \varpi_j(z_jI-\mathcal H)^{-1}.
\]
The coefficients include the contour orientation
and the factor $1/(2\pi i)$.
Because the underlying Gauss--Legendre weights
are positive and integrate constants exactly,  $\sum_{j=1}^m|\varpi_j|
 ={s_\Gamma}/{(2\pi)}
 \le{6\alpha_{\mathcal H}}/{\pi}$.

\subsubsection{Normalization and Error Allocation}
Each term in $S_m$ is a shifted inverse of
the Hamiltonian.
We implement these inverses using the uniform
bound $2M$, rather than requiring separate
estimates of their norms.
This is the conservative inverse-scale
implementation described in
\cite[Appendix~A.4 and Appendix~G.4,
Eqs.~(G.20)--(G.21)]{wang2026near}.
It permits a common inverse normalization
$O(M)$, including the constant-factor slack
needed for approximate Hamiltonian access.

The linear-combination construction therefore
has normalization bounded by
\[
 O(M)\sum_{j=1}^m|\varpi_j|
 =O(\alpha_{\mathcal H}M).
\]
Define
\[
 K:=\alpha_{\mathcal H}M=O(h^{12})
\]
and fix a known normalization
 $\alpha_\Pi:=c_\Pi K$, where $c_\Pi$ is a sufficiently large absolute
constant.
This normalization is independent of the
quadrature order and target accuracy.

Choose the projector tolerance
\[
 \eta_\Pi:=
 \frac{\varepsilon}{16(1+p_0^2)}.
\]
This is the tolerance required by the subsequent
Riccati-recovery step to achieve solution error
at most $\varepsilon/2$.
We divide the projector error budget equally
among quadrature error, Hamiltonian-input error,
and circuit implementation error.

First, choose the quadrature order prescribed
by Proposition~F.3 of \cite{wang2026near} so that
 $\|S_m-\Pi\|\le{\eta_\Pi}/{3}$. Second, use the Hamiltonian encoding constructed
above with error
 $\|\widehat{\mathcal H}-\mathcal H\|\le\xi$, where 
 $$
 \xi:=
 \min\left\{
 \frac1{4M},
 \frac{\eta_\Pi}
      {64\alpha_{\mathcal H}M^2}
 \right\}.
$$
To explain these two requirements, define the
quadrature sum with the approximate Hamiltonian:
\[
 \widehat S_m:=
 \sum_{j=1}^m
 \varpi_j(z_jI-\widehat{\mathcal H})^{-1}.
\]
The condition $\xi\le1/(4M)$ ensures that all
these shifted matrices remain invertible:
 $\xi\|(z_jI-\mathcal H)^{-1}\|
 \le2M\xi\le 1/2$.

The perturbation bound in
\cite[Proposition~A.5]{wang2026near},
with exact right factor $I$, then gives
\[
 \begin{aligned}
 \|\widehat S_m-S_m\|
 &\le
 \sum_{j=1}^m|\varpi_j|
 \frac{
 \xi\|(z_jI-\mathcal H)^{-1}\|^2
 }{
 1-\xi\|(z_jI-\mathcal H)^{-1}\|
 }\\
 &\le
 \frac{48}{\pi}
 \alpha_{\mathcal H}M^2\xi
 \le\frac{\eta_\Pi}{3}.
 \end{aligned}
\]
Thus the second requirement on $\xi$ makes
the Hamiltonian-input contribution fit within
its allocated budget; the constant $64$ is
simply a sufficient absolute constant.

Finally, implement the inverse and
linear-combination circuits accurately enough
that their represented output $\widehat\Pi$
satisfies
\(
 \|\widehat\Pi-\widehat S_m\|
 \le {\eta_\Pi}/{3}.
\)
Combining the three contributions yields
$ \|\widehat\Pi-\Pi\| \le
 \|\widehat\Pi-\widehat S_m\|
 +\|\widehat S_m-S_m\|
 +\|S_m-\Pi\|\le\eta_\Pi$.

\subsubsection{Cost of Projector Access}
Define
\[
 L:=\log\frac{2hK(1+p_0)}{\varepsilon}.
\]
For the chosen tolerances,
\[
 \log\frac{K}{\eta_\Pi}=O(L),
 \qquad
 \log\frac{2h}{\xi}=O(L).
\]
Proposition~F.3 of \cite{wang2026near} supplies a quadrature rule with
\[
 m=O(KL)
\]
nodes.
At each node, the shifted matrix has encoding
normalization $O(\alpha_{\mathcal H})$ and
inverse-norm bound $O(M)$.
The uniform inverse construction therefore uses
$O(KL)$ Hamiltonian-encoding calls per
projector call.
The node register is processed coherently:
each Hamiltonian call acts on the superposition
of nodes, so this query count is not multiplied
by $m$.

Each Hamiltonian-encoding call costs
\[
 Q_{\mathcal H}(\xi)
 =O\!\left(h^4\log\frac{2h}{\xi}\right)
 =O(h^4L)
\]
lookup-circuit evaluations.
Consequently, the implemented projector encoding
has normalization $ \alpha_\Pi=c_\Pi K$ and error
  $\|\widehat\Pi-\Pi\|\le\eta_\Pi$, and each call to it, its adjoint, or their
controlled versions requires at most
\[
 Q_\Pi(\eta_\Pi)
 =O(KL)\,Q_{\mathcal H}(\xi)
 =O(h^4KL^2)
\]
evaluations of the supplied lookup circuits.
Quadrature-node and coefficient preparation,
coherent arithmetic, and circuit synthesis
have polynomial additional cost.

\subsubsection{Recovering the Riccati Solution}
By
\cite[Lemma~3.6]{wang2026near},
the stable projector has range
$\operatorname{ran}[I;P_\star]$.
The direct recovery guarantees of
\cite[Proposition~D.5 and Appendix~F.1]
{wang2026near}
therefore apply with the supplied solution bound
$p_0$.

Choose the known rational output normalization  $\alpha_P:=8\alpha_\Pi(1+p_0)
 =O\!\left(K(1+p_0)\right)=O(h^{18})$. This satisfies
$\alpha_P\ge8\alpha_\Pi\sqrt{1+p_0^2}$.
For target Riccati error
$\varepsilon_P:=\varepsilon/2$,
our projector tolerance is precisely
\[
 \eta_\Pi
 =\frac{\varepsilon_P}{8(1+p_0^2)}.
\]
Use the inverse and product tolerances prescribed
in the cited recovery result:
\[
 \varepsilon_{\mathrm{inv}}
 =\frac{\varepsilon_P}{4(1+\alpha_\Pi)},
 \qquad
 \varepsilon_{\mathrm{prod}}
 \le\frac{\varepsilon_P}{4}.
\]
That result yields a solution encoding with
represented matrix $\widehat P_\star$ satisfying
  $\|\widehat P_\star-P_\star\|
 \le\varepsilon_P={\varepsilon}/{2}$.

The recovery uses
\[
 O\!\left(
 \alpha_P\log\left(e+\frac{\alpha_P}{\varepsilon}\right)
 \right)
\]
projector-encoding calls.
Hence one solution-encoding call, including its
adjoint or controlled version, requires
\[
 Q_P
 =O(h^4K\alpha_PL^3)
\]
lookup-circuit evaluations, with polynomial
additional gate and preprocessing costs.

\subsubsection{Estimating the Value}
The basis state $b=e_{i_b}$ is efficiently
preparable.
A Hadamard test on the solution encoding has
expectation
 $\mu
 ={\Re
 (b^\intercal\widehat P_\star b)}/{\alpha_P}$. Estimate $\mu$ to additive error
$\varepsilon/(2\alpha_P)$ and multiply the
result by $\alpha_P$.
Hoeffding's inequality gives the sufficient
repetition count
\[
 N=
 O\!\left(
 \frac{\alpha_P^2}{\varepsilon^2}
 \log\frac1\delta
 \right).
\]
With probability at least $1-\delta$, the
resulting estimate $\widehat V$ satisfies  $\left|
 \widehat V-
 \operatorname{Re}
 (b^\intercal\widehat P_\star b)
 \right|
 \le {\varepsilon}/{2}$. Since $\|b\|=1$ and $P_\star$ is real symmetric,
 $\left|
 \Re
 (b^\intercal\widehat P_\star b)
 -b^\intercal P_\star b
 \right|
 \le\|\widehat P_\star-P_\star\|
 \le {\varepsilon}/{2}$. Thus
 $|\widehat V-V(b)|\le\varepsilon$ with probability at least $1-\delta$.

The total lookup-query cost is bounded by
\[
 O\!\left(
 \frac{h^4K\alpha_P^3L^3}{\varepsilon^2}
 \log\frac1\delta
 \right).
\]
All other costs are polynomial in the encoded
input length, $h$, $1/\varepsilon$, and
$\log(1/\delta)$.
For fixed $D$, the quantity $h=(2+w)^D$ is
polynomial in the encoded instance length.
All precision choices, classical preprocessing,
and quantum circuit constructions are uniform.
This proves the stated estimation guarantee.

Finally, let $\Delta=a_1-a_0$ denote the promised
threshold gap.
For the decision problem, take
 $\delta=1/3$ and
 $\varepsilon=\min\{1,\Delta/4\}$, and compare $\widehat V$ with $(a_0+a_1)/2$.
On the successful estimation event this comparison
distinguishes the two promised cases.
Since $\Delta^{-1}\le h$, the algorithm runs in
polynomial time and succeeds with probability
at least $2/3$.
Therefore $\textsc{SLQRV}\in\BQP$.


\section{Completeness Results}
The embedding in \cref{prop:Embedding}, together with the
hardness result in \cref{thm:main} and the membership result
in \cref{thm:lqrv-membership}, establishes the following.

\begin{corollary}
\label{thm:slv-complete}
For every fixed pair of positive integers $s,d$,
$\textsc{SLV}$ belongs to $\BQP$ and is $\BQP$-complete
when $s=d=5$.
Furthermore, $\textsc{SLQRV}$ is $\BQP$-complete
for every sufficiently large fixed integer $D$.
Both completeness results hold under classical
polynomial-time Karp reductions.
\end{corollary}

These results establish a conditional worst-case advantage
of quantum algorithms for infinite-horizon performance
evaluation. Both problems admit bounded-error quantum
algorithms running in time polynomial in the succinct input
length. Unless $\BPP=\BQP$, no classical randomized algorithm
can achieve the same polynomial-time guarantee on all
promised instances: such an algorithm, composed with the
hardness reductions, would efficiently simulate every
$\BQP$ computation. This comparison uses the same succinct
input representation and scalar decision task for both
computational models. The separation concerns worst-case
complexity; it neither establishes a specific exponential
speedup nor excludes efficient classical algorithms for
restricted classes of systems.

\section{Discussion}
\label{sec:Discussion}
\subsection{Control Interpretation of the Hard Instances}
Recall \cref{thm:main} and its proof. 
For each input length \(r\), the reduction uses the same \(A_r\), \(\alpha_L\), and \(\gamma_{L,\mathrm r}\); the input \(x\) appears only in \(b_x\) and \(c_x\). The clock operator satisfies \(\bra{c_x}W_r^L\ket{b_x}=p_x\), embedding the circuit computation in a terminal-clock matrix element.

Every eigenvalue of \(A_r\) is \(-\alpha_L\), where \(\alpha_L=\Theta(L^{-1/2})\), so the hard instances are strictly Hurwitz. Their exact dissipativity margin is \(\gamma_L=\Theta(L^{-5/2})\), and the rational certificate \(\gamma_{L,\mathrm r}\) has the same order. Since \(L=\Theta(\log n)\), \(\gamma_{L,\mathrm r}=\Theta((\log n)^{-5/2})\) and \(\|\exp(A_r t)\|\le\exp(-\gamma_{L,\mathrm r}t)\). The hardness arises from the nonnormal feed-forward dynamics (implied in the computation of $W_{r}^{L}$): a normal matrix \(A=-\alpha_L I\) with the same eigenvalue has the readily computable value \(J=(c^\intercal b)^2/(2\alpha_L)\), where \(c^\intercal b\in\{0,1\}\).

\subsection{A Uniformly Positive Dissipativity Margin Gives an Easy Classical Regime}
A shrinking margin prevents a simple classical truncation method from applying to the hard instances. Consider SLV instances with \(-(A+A^\intercal)/{2}\succeq\gamma I\), \(\|A\|\le1\), and coordinate vectors \(b,c\). For any \(T\ge0\), the tail satisfies $\int_T^\infty |c^\intercal e^{At}b|^2\,\mathrm{d}t\le {e^{-2\gamma T}}/(2\gamma)$.
Thus, if \(\gamma\ge\gamma_0>0\) and the promise gap is constant, a constant \(T\) makes this tail smaller than a fixed fraction of the gap. On \([0,T]\), approximate \(e^{At}b\) by $q_K(t)=\sum_{k=0}^K\frac{t^k}{k!}A^kb$. Since \(\|A\|\le1\), 
\[\|e^{At}-\sum_{k=0}^K\frac{t^k}{k!}A^k\|\le e^T\frac{T^{K+1}}{(K+1)!},~t\in \intcc{0,T}.\]
For constant \(T\), constant \(K\) suffices for constant accuracy. If each column of \(A\) has at most \(s\) nonzeros, \(A^kb\) has at most \(s^k\) nonzeros, so the required coefficients take \(s^{O(K)}\) sparse-entry lookups, up to polynomial factors in the input length. This gives a deterministic polynomial-time algorithm when the dissipativity margin is uniformly positive.

The hard family instead has \(\gamma_{L,\mathrm{r}}^{-1}=\Theta(L^{5/2})\). Moreover, \(c_x^\intercal A_r^k b_x=0\) for \(k<L\), while \(c_x^\intercal A_r^L b_x=\alpha_L^L p_x\). A truncation below degree \(L\) therefore misses the encoded acceptance probability. The shrinking margin rules out this constant-degree truncation approach; the hardness itself comes from the \(\BQP\)-hard circuit evaluation embedded in the terminal-clock contribution.

\subsection{Establishing $\BQP$-Hardness for the Controlled Case}
The clock family in the proof of \cref{thm:main} also gives hardness when control acts on the measured
output, without a second circuit construction. Use its padded drift
$A_r$, terminal coordinate $c$, and initial coordinate $b$, and take
$B=(\alpha_L/4)c$, $C=c^{\intercal}$, and $R=[1]$. The terminal clock
has no successor, so $A_r c=-\alpha_L c$. The impulse response from
$u$ to the output is $g(t)=(\alpha_L/4)e^{-\alpha_L t}$, and causal
convolution $T_g$ has $L^2$ operator norm at most $\|g\|_{L^1}=1/4$.
For $f(t)=c^{\intercal}e^{A_r t}b$, the controlled output is $f+T_g u$.
The triangle and Cauchy--Schwarz inequalities give
 $\|f\|_{L^2}^2\le(1+\|T_g\|^2)
       (\|f+T_g u\|_{L^2}^2+\|u\|_{L^2}^2)$.
Taking the infimum, and using $u=0$ for the opposite inequality, gives
$(16/17)p_x^2\le V(b)\le p_x^2$, since Section~\ref{sec:Hardness}
proved $\|f\|_{L^2}^2=p_x^2$. On NO instances, $V(b)\le4/25$;
on YES instances, $V(b)\ge144/425$. The additional actuator is
dyadic and has a one-entry description. The drift and the zero-gain
stability witnesses are unchanged. Hence the hardness result also holds
for this single-input, single-output restriction with $CB\ne0$.

\section{Conclusion}
\label{sec:Conclusion}
We have established that the infinite-horizon succinct Lyapunov and 
LQR value problems are $\BQP$-complete under their respective promises.
Both problems admit quantum algorithms running in time polynomial
in the succinct input length, whereas a classical randomized
algorithm with the same worst-case guarantee would imply
$\BPP=\BQP$. These results identify a conditional quantum advantage
for evaluating fundamental infinite-horizon control-performance
measures under a common input representation. Future work may explore extensions to other input and output models
and identify structured system classes that admit more efficient
classical or quantum evaluation algorithms.
\section*{Acknowledgment}
The authors thank Maxwell Fitzsimmons and Meenu Kumari for valuable
discussions and feedback on this work.
ChatGPT~\cite{chatgpt} and OpenAI Codex~\cite{codex} were used to search
the literature discussed in Section~\ref{sec:Introduction}, to check algebraic
steps in the proofs in Sections~\ref{sec:Hardness} and~\ref{sec:Membership}
and Appendices~\ref{app:Circuits}--\ref{sec:Proof_lem:lqr-conditioning}, and to
edit the grammar and syntax of the text in all sections. The authors verified
all proofs and take full responsibility for the content of the paper.

\appendices
\crefalias{section}{appendix}
\section{Proof of Lemma~\ref{thm:ComplexCircuitRationalApproximation}}
\label{app:Circuits}
The proof has three steps. First, we show that complex circuits can be
simulated by real circuits (\cref{lem:real-simulation}). Second, we show
that real circuits can be approximated arbitrarily closely using Hadamard
and Toffoli gates (\cref{lem:HT-universality}). Finally, we replace each
Hadamard gate with a circuit whose gate entries are rational, at the cost
of enlarging the register (\cref{lem:dyadic-gate-replacement}).

\begin{lemma}[Real simulation of quantum circuits]
\label{lem:real-simulation}
Let $\{Y_r\}_{r\ge1}$ be a polynomial-time uniform
family of quantum circuits over a fixed finite
set $\mathcal G$ of complex unitary gates,
acting on $q(r)$ qubits with $\ell(r)$ gates,
where $q(r),\ell(r)$ are polynomially bounded.

There is a polynomial-time uniform family
$\{\widetilde Y_r\}_{r\ge1}$ over a fixed finite
set of real orthogonal gates, acting on
$q(r)+1$ qubits with exactly $\ell(r)$ gates,
that preserves every acceptance probability.
The additional qubit is initialized to $\ket0$
and ignored by the acceptance measurement.

Consequently, every problem in $\BQP$ admits
real quantum circuits with the same bounded-error
completeness and soundness parameters.
\end{lemma}

\begin{proof}
Following \cite{aharonov2003,bernstein1993quantum},
for $U=A+iB\in\mathbb C^{d\times d}$ define
\begin{equation}
 \mathsf R(U)
 \defas
 \begin{bmatrix}
  A&-B\\
  B&A
 \end{bmatrix}.
 \label{eq:realification}
\end{equation}
Direct multiplication gives
 $\mathsf R(UV)=\mathsf R(U)\mathsf R(V)$ and
 $\mathsf R(U^\dagger)=\mathsf R(U)^\intercal$. Hence, if $U$ is unitary, then
 $\mathsf R(U)^\intercal\mathsf R(U)
 =\mathsf R(U^\dagger U)=I_{2d}$, so $\mathsf R(U)$ is real orthogonal.

For $\ket{\psi}=u+iv$, with $u,v\in\mathbb R^d$,
define  $\mathcal E(\ket{\psi})
 :=\ket0_a\otimes u+\ket1_a\otimes v$.
Then
 $\mathsf R(U)\mathcal E(\ket{\psi})
 =\mathcal E(U\ket{\psi})$. Construct $\widetilde Y_r$ by replacing each
gate $G$ of $Y_r$ with $\mathsf R(G)$, acting
on its original qubits and the shared additional
qubit $a$.
The preceding identity shows that the resulting
circuit implements $\mathsf R(U_r)$, where
$U_r$ is the unitary implemented by $Y_r$.

For input $x$, let $\ket{\psi_x}$ be the
original computational-basis input, including
any zero-initialized ancillas, and write
$U_r\ket{\psi_x}=u_x+iv_x$.
Since $\ket{\psi_x}$ is real,
 $\widetilde Y_r
 (\ket0_a\otimes\ket{\psi_x})
 =\ket0_a\otimes u_x+\ket1_a\otimes v_x$. For the original computational-basis acceptance
projector $P_{\mathrm{acc}}$, the simulator's
acceptance probability is therefore
 $\|P_{\mathrm{acc}}u_x\|^2
 +\|P_{\mathrm{acc}}v_x\|^2
 =\|P_{\mathrm{acc}}U_r\ket{\psi_x}\|^2$,
as required.

Finally, $\mathsf R(\mathcal G)$ is a fixed
finite gate set of bounded arity.
The construction adds one qubit and replaces
each original gate by one real gate.
A uniformity algorithm generates $Y_r$ and
performs these fixed gate-label and support
replacements with polynomial overhead.
Thus $\{\widetilde Y_r\}$ is polynomial-time
uniform and has exactly $\ell(r)$ gates.
\end{proof}

\begin{lemma}[Hadamard and Toffoli gates \cite{shi2003,aharonov2003}]
\label{lem:HT-universality}
Let \(\mathcal{F}\) be a fixed finite set of real orthogonal gates, each
acting on at most a constant number of qubits. Let
\(\{Y_r\}_{r\geq 1}\) be a polynomial-time uniform family of quantum
circuits over \(\mathcal{F}\), where \(Y_r\) acts on \(q(r)\) qubits and has
at most \(\ell(r)\) gates, for polynomials \(q\) and \(\ell\). Then, for every polynomial \(p:\mathbb{N}\to\mathbb{N}\), there exists a
polynomial-time uniform family of circuits $\{C_r\}_{r\geq 1}$ composed solely of Hadamard and Toffoli gates, acting on
\(q(r)+a(r)\) qubits for a polynomially bounded function \(a\), such that,
for every \(r\geq1\),
\begin{equation}
    \left\|
    C_r\bigl(\,\cdot\,\otimes\ket{\mathbf{0}}\bigr)
    -(Y_r\,\cdot\,)\otimes\ket{\mathbf{0}}
    \right\|
    \leq \frac{1}{p(r)}.
    \label{eq:HT-approximation}
\end{equation}
Here, 
\(C_r(\,\cdot\,\otimes\ket{\mathbf{0}})\) is regarded as a map from the
original \(q(r)\)-qubit state space into the enlarged state space. In particular, if \(Y_r\) accepts an input \(x\in\{0,1\}^r\) with
probability \(p_{Y_r}(x)\), and \(C_r\) accepts according to the same output
qubit, then  
\[
\bigl|p_{C_r}(x)-p_{Y_r}(x)\bigr|
    \leq \frac{2}{p(r)}.
\] Consequently, if \(\{Y_r\}_{r\geq1}\) decides a promise problem with a
constant bounded-error gap, then \(p\) can be chosen so that
\(\{C_r\}_{r\geq1}\) decides the same promise problem with a constant
bounded-error gap.
\end{lemma}

\begin{lemma}
\label{lem:dyadic-gate-replacement}
Let
\(
    C=G_TG_{T-1}\cdots G_1
\)
be a real quantum circuit on \(q\) qubits, where each \(G_t\) is either a
Hadamard gate or a Toffoli gate. Let \(h\) denote the number of Hadamard
gates in this decomposition. Then, one can construct a real quantum circuit
\(
    \widetilde C
    =
    \widetilde G_T\widetilde G_{T-1}\cdots\widetilde G_1
\)
on $Q\defas q+h$
qubits with the following properties:
\begin{enumerate}
    \item Each component matrix \(\widetilde G_t\) is real orthogonal and
    has all of its entries in
    \(
        \left\{0,\pm 1/2,\pm 1\right\}.
    \)

    \item For every state \(\ket{\phi}\) of the original \(q\)-qubit
    register,
    \begin{equation}
        \widetilde C
        \bigl(\ket{\phi}\otimes\ket{0}^{\otimes h}\bigr)
        =
        \bigl(C\ket{\phi}\bigr)\otimes\ket{+}^{\otimes h}.
        \label{eq:dyadic-replacement-action}
    \end{equation}

    \item If \(P\) is an orthogonal projector acting on the original
    \(q\)-qubit register, then, with
    \(
        \widetilde P\defas P\otimes I_{2^h},
    \)
    one has
    \begin{equation}
        \left\|
            \widetilde P\,
            \widetilde C
            \bigl(\ket{\phi}\otimes\ket{0}^{\otimes h}\bigr)
        \right\|^2
        =
        \left\|PC\ket{\phi}\right\|^2.
        \label{eq:dyadic-replacement-probability}
    \end{equation}
\end{enumerate}
Moreover, the construction has polynomial overhead: it adds \(h\leq T\)
qubits, does not increase the number of circuit components, and preserves
polynomial-time uniformity.
\end{lemma}
\begin{proof}
Enumerate the Hadamard gates in the ordered decomposition of \(C\). For
each \(j\in \intcc{1;h}\), introduce a fresh ancillary qubit \(a_j\),
initialized in the state \(\ket{0}\).
Suppose that \(G_t\) is the \(j\)-th Hadamard gate and that it acts on
original-register qubit \(u\). Define
\(
    \widetilde G_t
    \defas
    H_u^{(Q)}H_{a_j}^{(Q)},
\)
where \(H_v^{(Q)}\) denotes a Hadamard gate acting on qubit \(v\) of the
enlarged \(Q\)-qubit register. Since the two Hadamard gates act on distinct
qubits, they commute. Up to a permutation of tensor factors,
\(
    \widetilde G_t
    =
    (H\otimes H)\otimes I_{2^{Q-2}}.
\)
Thus every entry of \(\widetilde G_t\) belongs to
\(
    \left\{0,\pm 1/2\right\}.
\)
If \(G_t\) is a Toffoli gate, let \(\widetilde G_t\) be the same Toffoli
gate acting on the original-register qubits and acting as the identity on
all fresh ancillary qubits. Then \(\widetilde G_t\) is a permutation matrix,
so all of its entries belong to \(\{0,1\}\). Therefore, in both cases,
\(
    (\widetilde G_t)_{ij}
    \in
    \left\{0, \pm 1/2,\pm 1\right\}
\)
for all matrix indices \(i,j\). Moreover, each \(\widetilde G_t\) is real
orthogonal. Each ancillary qubit \(a_j\) is used only in the replacement of the \(j\)-th
Hadamard gate. Hence, immediately before that replacement, \(a_j\) is still
in the state \(\ket{0}\). If the state of the original register immediately
before that gate is \(\ket{\varphi}\), then
$
    \widetilde G_t
    \bigl(\ket{\varphi}\otimes\ket{0}_{a_j}\bigr)
    =
    (H_u\ket{\varphi})\otimes H\ket{0}_{a_j}=
    (H_u\ket{\varphi})\otimes\ket{+}_{a_j}.
$
No subsequent component acts on \(a_j\). The Toffoli replacements act
exactly as the original Toffoli gates on the original register and leave all
fresh ancillary qubits unchanged. An induction over
\(t\in \intcc{1;T}\) consequently gives
\(
    \widetilde C
    \bigl(\ket{\phi}\otimes\ket{0}^{\otimes h}\bigr)
    =
    \bigl(C\ket{\phi}\bigr)\otimes\ket{+}^{\otimes h},
\)
which proves the second item \eqref{eq:dyadic-replacement-action}. Finally, using \eqref{eq:dyadic-replacement-action} and
\(\|\ket{+}^{\otimes h}\|=1\), we obtain
    $\left\|
        \widetilde P\,
        \widetilde C
        \bigl(\ket{\phi}\otimes\ket{0}^{\otimes h}\bigr)
    \right\|^2=
    \left\|PC\ket{\phi}\right\|^2$,
proving \eqref{eq:dyadic-replacement-probability}. The construction adds one fresh qubit for each Hadamard gate, so
\(Q=q+h\leq q+T\), and replaces each original circuit component by exactly
one new component. Given a description of \(C\), the description of
\(\widetilde C\) can be generated by scanning the ordered gate list once and
assigning a fresh ancillary qubit to each Hadamard gate. Thus the
construction preserves polynomial circuit size and polynomial-time
uniformity.
\end{proof}

\begin{proof}[Proof of \cref{thm:ComplexCircuitRationalApproximation}]
By \cref{lem:real-simulation}, each gate of \(Y_r\) can be replaced by a
real orthogonal gate acting on one additional qubit, while preserving the
acceptance probability exactly. By \cref{lem:HT-universality}, the resulting
real circuit can be approximated by a polynomial-time uniform
Hadamard--Toffoli circuit to operator-norm error at most \(1/p(r)\). Next, by \cref{lem:dyadic-gate-replacement}, replace each Hadamard gate by
its two-qubit implementation \(H\otimes H\), with a fresh ancilla initialized
to \(\ket{0}\). This replacement preserves acceptance probabilities and has
rational matrix entries in \(\{0,\pm {1}/{2},\pm 1\}\). The resulting
family is polynomial-time uniform and has polynomial size. Its acceptance
probability differs from that of \(Y_r\) by at most \(2/p(r)\), as claimed.
\end{proof}


\section{Spectrum of the Clocked Propagation Operator}
\label{app:clock-spectrum}
Recall that
    $W_r
    =
    \sum_{k=0}^{L-1}
    \ket{k+1}_{\rm clk}\!\bra{k}_{\rm clk}
    \otimes G_{r,k+1}$,
where each \(G_{r,k}\in\mathbb{R}^{2^L\times2^L}\) is real orthogonal. For \(k\in\intcc{0;L}\), define
    $P_0\defas I_{2^L}$,     $P_k\defas G_{r,k}\cdots G_{r,1}$,  
    $k\in \intcc{1;L}$, and let $R  \defas
    \sum_{k=0}^{L}
    \ket{k}_{\rm clk}\!\bra{k}_{\rm clk}\otimes P_k$. Since every \(P_k\) is real orthogonal, \(R\) is real orthogonal. Moreover,
the identity
    $G_{r,k+1}P_k=P_{k+1}$, for $k\in \intcc{0;L-1}$, 
implies that
   $R^{\intercal}W_rR
    =
    \sum_{k=0}^{L-1}
    \ket{k+1}_{\rm clk}\!\bra{k}_{\rm clk}
    \otimes I_{2^L}$.
Hence, the following similarity is established:
\begin{equation}
    R^{\intercal}\left(\frac{W_r+W_r^\intercal}{2}\right)R
    =
    J_L\otimes I_{2^L},
    \label{eq:clock-orthogonal-similarity}
\end{equation}
where 
\[
    J_L
    \defas
    \frac{1}{2}
    \sum_{k=0}^{L-1}
    \left(
        \ket{k+1}_{\rm clk}\!\bra{k}_{\rm clk}
        +
        \ket{k}_{\rm clk}\!\bra{k+1}_{\rm clk}
    \right).
\]
For \(j\in \intcc{1;L+1}\), define
$\theta_j\defas{j\pi}/(L+2)$
and
\[
    \ket{u^{(j)}}
    \defas
    \sum_{k=0}^{L}
    \sin\bigl((k+1)\theta_j\bigr)\ket{k}_{\rm clk}.
\]
Using
$    \sin(k\theta)+\sin((k+2)\theta)
    =
    2\cos(\theta)\sin((k+1)\theta)$, 
together with the boundary identities
    $\sin(0)=0$ and $\sin\bigl((L+2)\theta_j\bigr)=\sin(j\pi)=0$,
we obtain
   $ J_L\ket{u^{(j)}}
    =
    \cos(\theta_j)\ket{u^{(j)}}$. The \(L+1\) eigenvalues
    $\cos\!\left({j\pi}/(L+2)\right)$,    $j\in \intcc{1;L+1}$, are distinct. Thus, they constitute the complete spectrum of \(J_L\). The similarity with $J_{L}\otimes I_{2^L}$ implies that the spectrum of $(W_r+W_r^{\intercal})/{2}$ is given by
\[
    \left\{
        \cos\!\left(\frac{j\pi}{L+2}\right)
        :
        j\in \intcc{1;L+1}
    \right\},
\]
with each eigenvalue repeated \(2^L\) times.

\section{Continuation of the Proof of Theorem~\ref{thm:main}}
\label{sec:ProofContinuation}
\subsection{Construction of the Row-Lookup Circuit}
Set $q_{\rm clk}\defas\lceil\log_2(L+1)\rceil$
and $\ell=q_{\rm clk}+L$.
Interpret each $\ell$-bit address as
$(\kappa,z)\in
\intcc{0;2^{q_{\rm clk}}-1}\times
\intcc{0;2^L-1}$, with one-based matrix index
$\iota(\kappa,z)\defas1+\kappa2^L+z$.
Clock addresses $0\le\kappa\le L$ belong to the
original block; larger addresses are padding.

We use five row slots.
For row $(\kappa,z)$, slot $1$ returns the diagonal
entry
$\overline A_r(\iota(\kappa,z),\iota(\kappa,z))
=-\alpha_L$.
If $1\le\kappa\le L$, the only off-diagonal block
is $\alpha_LG_{r,\kappa}$ at column-clock position
$\kappa-1$, so
$\overline A_r(\iota(\kappa,z),\iota(\kappa-1,y))
=\alpha_L(G_{r,\kappa})_{z,y}$.
Slots $2,\ldots,5$ enumerate the at most four
nonzero entries in row $z$ of $G_{r,\kappa}$:
an entry $(G_{r,\kappa})_{z,y}=g\ne0$ produces
$(\iota(\kappa-1,y),\alpha_Lg)$.
Any remaining slots return zero with the
prescribed out-of-range column index from
Section~\ref{sec:ProblemFormulation}-\ref{subsec:sparse-input}.
For $\kappa=0$ or $\kappa>L$, all four
off-diagonal slots are unused.
Thus every row has at most five nonzero entries.

The reduction stores the ordered component
descriptions $G_{r,1},\ldots,G_{r,L}$, each
specifying its type and $O(1)$ acted-on qubits
in $O(\log L)$ bits.
A multiplexer selects $G_{r,\kappa}$ using
$O(L\log L)$ Boolean gates.
With $O(L)$ additional gates, the circuit
enumerates its local nonzero row entries,
constructs $y$ by replacing the acted-on bits
of $z$, and attaches the exact dyadic value
$\alpha_Lg$, where
$g\in\{0,\pm1/2,\pm1\}$.
Boundary checks and unused-slot outputs also
require $O(L)$ gates.

Hence the row-lookup circuit has size $O(L^2)$.
Its output uses the column-index width prescribed
in Section~\ref{sec:ProblemFormulation}-\ref{subsec:sparse-input}, including
room for dummy indices, and a dyadic word whose
numerator and denominator exponent each have
$O(L)$ bits.
The total output length is therefore $O(L)$,
including unused-slot outputs.
\subsection{Column Lookup and Verification of the SLV Promises}
The column-lookup circuit is analogous. For a column indexed by
\((\kappa,z)\), its first slot again specifies the diagonal entry
\(-\alpha_L\). If \(0\leq\kappa\leq L-1\), its possible off-diagonal
entries lie in the block joining clock position \(\kappa\) to clock position
\(\kappa+1\). More precisely,
    $\overline A_r\bigl(
        \iota(\kappa+1,y),\iota(\kappa,z)
    \bigr)
    =
    \alpha_L(G_{r,\kappa+1})_{y,z}$.

Since every column of \(G_{r,\kappa+1}\) has at most four nonzero entries,
the remaining four slots enumerate these entries. For \(\kappa=L\) and for
padded clock addresses, all four off-diagonal slots are unused.

The column-lookup circuit has the same size and output-length bounds.
Because  $\ell=\Theta(L)$,
both lookup circuits have size \(O(\ell^2)\), and every output word has
\(O(\ell)\) bits.  From the explicit representation above,
\(\gamma_{L,\mathrm{r}}\) has an \(O(\ell)\)-bit numerator--denominator
encoding.  In addition, the preceding inverse bound and
\(\ell\geq L+1\) give
    $\gamma_{L,\mathrm{r}}^{-1}
    \leq
    2\sqrt{L}\,(L+2)^2
    \leq
    \ell^5$. Moreover, since $\Delta=a_1-a_0=1/5$ and $d=5$, we have
$\Delta^{-1}=5\le \ell(r)^d$ for $\ell(r)\ge2$, as ensured by the padding.
An $O(\ell^2)$-gate circuit has an
$O(\ell^2\log(2+\ell))$-bit gate-list description. The fixed padding
constant $L_{\min}$ makes this bound and the $O(\ell)$ word bounds at most
$\ell^5$ for every output. Thus the fixed choice \(d=5\) satisfies every size, output-length, and
dissipativity-inverse promise in the definition of \textsc{SLV}.
\subsection{Polynomial-Time Construction of the Reduction}
Finally, define the output data of the reduction by
    $A_r\defas\overline A_r$,
    $\gamma_r\defas\gamma_{L,\mathrm{r}}$, 
    and
    $\alpha_r\defas\alpha_L$.
 The vectors \(b\) and \(c\) are computational-basis vectors, so their
indices \(i_b(x)\) and \(i_c(x)\) are obtained directly from the clock--work
addresses in \eqref{eq:bc} and can be computed in time polynomial in \(r\).
The matrix \(A_r\) and the rational number \(\gamma_r\) depend only on
\(r\) and on the fixed circuit family for \(\Pi\), while these two indices
are the only output data that depend on \(x\).  The construction of the
gate list, the two lookup circuits, the rational encodings, and the indices
uses time polynomial in \(r\); it never forms the \(n\times n\) matrix
\(\overline A_r\).   This completes the
reduction and the proof of Theorem~\ref{thm:main}.
\section{Proof of Lemma~\ref{lem:lqr-conditioning}}
\label{sec:Proof_lem:lqr-conditioning}
For any initial state $v$, the comparison policy $u=-Fx$ gives
\begin{align*}
 v^{\intercal}P_\star v
 &\le (\|C\|^2+\|R\|\,\|F\|^2)
       \int_0^\infty h^2e^{-2t/h}\,\mathrm{d}t\,\|v\|^2\\
 &\le \tfrac12 h^3(h^2+h^3)\|v\|^2\le h^6\|v\|^2.
\end{align*}
Since $P_\star \succeq0$, this proves \eqref{eq:lqr-P-bound}.

For the resolvent estimate, put $B_0=BR^{-1/2}$ and $F_0=R^{1/2}F$.
Then $B_0B_0^{\intercal}=G$, $B_0F_0=BF$, and
$\max\{\|B_0\|,\|F_0\|,\|C\|,\|L\|\}\le h^{3/2}$. These factors
are used only in the proof. Fix $\omega\in\mathbb R$, let
$D_\omega=i\omega I-A$, and write
$(f,g)=(i\omega I-\mathcal H)(x,y)$ for complex $x,y\in\mathbb C^n$.
Then $f=D_\omega x+Gy$ and $g=Qx-D_\omega^\dagger y$. Cancellation
of the cross terms in the real part gives $E\defas\|Cx\|^2+\|B_0^{\intercal}y\|^2=\Re(y^\dagger f+x^\dagger g)\le ZW$, where $Z=\|(x,y)\|$ and $W=\|(f,g)\|$. By the Laplace representation
of the resolvent and \eqref{eq:lqr-control-promise}, both
$(D_\omega+LC)^{-1}$ and $(D_\omega+BF)^{-1}$ have norm at most
$r\defas\int_0^\infty h e^{-t/h}\,\mathrm{d}t=h^2$. The same holds
for their adjoints. Apply these bounds to
\begin{align*}
 (D_\omega+LC)x&=f-B_0(B_0^{\intercal}y)+L(Cx),\\
 (D_\omega+BF)^\dagger y&=-g+C^{\intercal}(Cx)+F_0^{\intercal}(B_0^{\intercal}y).
\end{align*}
The two correction terms in each row have combined norm at most
$\sqrt{2}h^{3/2}\sqrt E$. Taking the norm of the pair therefore gives
$Z\le rW+2rh^{3/2}\sqrt{ZW}$. Since the last term is at most
$Z/2+2r^2h^3 W$, we get $Z\le(2r+4r^2h^3)W$. This proves invertibility
in finite dimensions and yields \eqref{eq:lqr-axis-bound} uniformly
in $\omega$.

\end{document}